%% file: main.tex
\documentclass[12pt]{article}

\usepackage{amsmath,amsfonts,amssymb,amsthm,booktabs,color,epsfig,graphicx,hyperref,url, setspace, algorithm, algorithmic, bm, bbm, natbib, setspace, arydshln, authblk, multicol, tkz-graph, rotating, tablefootnote}
\usepackage[margin=1in]{geometry}
\usepackage[title]{appendix}
\usetikzlibrary{shapes, arrows, calc, decorations.pathreplacing}
\newcommand{\parheader}[1]{\vskip12pt\noindent\textbf{#1.}}
\DeclareMathOperator{\argmin}{argmin}

\begin{document}

\title{\bf Sparse estimation of parameter support sets for generalized vector autoregressions by resampling and model aggregation}

\author[1]{Trevor D. Ruiz\thanks{Corresponding Author: truiz01@calpoly.edu}}
\affil[1]{
    Statistics Department, California Polytechnic State University, San Luis Obispo, CA}
\author[2]{Sharmodeep Bhattacharyya}
\author[2]{ and Sarah C. Emerson}
\affil[2]{
    Department of Statistics,
    Oregon State University,
    Corvallis, OR
    }

\date{}

\maketitle
\bigskip
\begin{abstract}
The central problem we address in this work is estimation of the parameter support set $S$, the set of indices corresponding to nonzero parameters, in the context of a sparse parametric likelihood model for discrete multivariate time series. We develop an algorithm that performs the estimation by aggregating support sets obtained by applying the LASSO to data subsamples. Our approach is to identify several candidate models and estimate $S$ by selecting common parameters, thus ``aggregating" candidate models. While our method is broadly applicable to any selection problem, we focus on the generalized vector autoregressive (GVAR) model class, and particularly the Poisson case, emphasizing applications in network recovery from discrete multivariate time series. We propose benchmark methods based on the LASSO, develop simulation strategies for GVAR processes, and present empirical results demonstrating the superior performance of our method. Additionally, we present an application estimating ecological interaction networks from paleoclimatology data.
\end{abstract}

\vfill

\newpage
\spacing{1.5}

\section{Introduction}\label{sec:intro}

Generalized vector autoregressive (GVAR) models are discrete-time stochastic processes $\{X_t\}_{t \in \mathbb{Z}}$ specified through a parametric model for $X_t|\mathcal{H}_t$ in the exponential family in which the parameter is a function of the process history $\mathcal{H}_t$. They are an extension of the vector autoregressive (VAR) model framework to non-Gaussian and discrete data \citep{cox1981statistical, fahrmeir1994multivariate, karlis2016models, fokianos2024multivariate}. VAR models can be viewed as special cases of GVAR models for Gaussian data, in much the same way that the normal linear model is a special case of the generalized linear model. Similarly, VAR models have a longer history of use. Both GVAR and VAR models describe the serial dependence in multivariate time series, but are densely parametrized: the number of parameters scales quadratically with the process dimension. For processes of even modest dimension, the dependence structure can be quite sparse relative to the number of model parameters; this makes interpretation of parameter estimates challenging for large datasets.

The LASSO \citep{tibshirani1996regression} is used to estimate sparse models in the VAR case \citep{bach2004learning, arnold2007temporal, hsu2008subset, basu2015regularized} and also in the GVAR case \citep{mark2017network, hall2016inferring, hall2018learning, pandit2020generalized, han2023high}. However, \citet{arnold2007temporal} observed that this approach tends to produce overfitting. Variants on the LASSO specifically adapted to the VAR setting sought to address this difficulty \citet{shojaie2010discovering, song2011large, fan2011sparse}, as well as alternative sparse estimation methods \citet{han2015direct, davis2016sparse}. However, few such alternatives exist for GVAR models. While LASSO estimates exhibit consistency and support recovery properties under certain conditions \citep{pandit2020generalized, han2023high}, these conditions are difficult to verify in practice and there is little work exploring the empirical performance of this approach.

Applications of VAR and GVAR models in the sciences have been around for some time \citep{brown2004multiple, valdes2005estimating, pillow2008spatio, paul2008multivariate, bracher2022endemic}, and a central theme in estimating VAR and GVAR models with sparsity is learning causal relationships \citep{dahlhaus2003causality, valdes2005estimating, lozano2009grouped, mark2017network, hall2018learning}. These relationships can be inferred directly from the sparsity pattern of parameter estimates: model parameters indicate the dependence structure among component time series. The parameter support set itself is thus an object of particular interest. Consequently, the selection behavior of sparse estimators in practice with limited data is of special importance.

In contexts besides time series analysis, data resampling has been used to formulate selection and sparse estimation methods that achieve strong empirical performance. Resampling allows for the possibility of obtaining collections of models with comparable sparsity levels that can be considered as candidates and leveraged in various ways to formulate more robust strategies for searching large model spaces. For example, if many resamples are considered, candidate models can be used to estimate the frequency with which parameters are selected across data subsets at different sparsity levels. \citet{bach2008bolasso} used these frequencies to reduce overfitting by returning only common parameter supports across all resamples. \citep{meinshausen2010stability} performed selection without fixing regularization hyperparameters by choosing parameters having a maximum selection frequency above a certain threshold. \citet{ruiz2020sparse} considered pooling supports meeting optimality criteria across resamples; \citet{capanu2023subsampling} proposed a similar approach for generalized linear models.

In this work we propose a sparse estimation method for generalized vector autoregressions based on simple aggregation operations applied to multiple estimates obtained from data resampling, and demonstrate the superiority of the approach compared with simpler alternatives. The method is developed in the context of first-order Poisson GVAR processes, but generalizes easily to any support set estimation or model selection problem. A high-level presentation of the methodology is given in Section \ref{sec:methods}. We provide empirical support for the method in Section \ref{sec:simulations} along with a discussion of special challenges in selecting parameter values for simulating process realizations. We give a data application in \ref{sec:application} illustrating the use of the methodology and its advantage over alternatives in practice. Section \ref{sec:discussion} closes the paper with a summary of findings.

\section{Methods}\label{sec:methods}

\subsection{Poisson GVAR processes}

We denote by $X_t$ an $M$-dimensional random vector indexed by time $t$; the $m$th element of $X_t$ is $X_{t, m}$. An $M$-dimensional stochastic process $\{X_t\}_{t \in \mathbb{Z}}$ is a generalized vector autoregressive (GVAR) process if the elements of $X_t$ are mutually conditionally independent given the process history with conditional distributions $p(x_{tm}; \theta_{tm})$ that depend on the past values through a parametric function $\theta_m(x_{t - 1}, x_{t - 2}, ...)$, \textit{i.e.}, if for every $t \in \mathbb{Z}$: 
\begin{equation*}
(X_{t, m}\;|\;X_{s} = x_s, s < t)
\stackrel{indep.}{\sim} p\left(\cdot\;; \theta_{t, m} = \theta_m (x_{t - 1}, x_{t - 2}, \dots)\right) 
\text{, }
m \in \{1, \dots, M\}
\end{equation*}
Given an initial distribution $\nu_0$, the joint distribution of any finite-time process realization $\{x_t\}_{t = 0}^T$ is:
\begin{equation*}
\nu_0 (x_0) \prod_{t = 1}^T \prod_{m = 1}^M p(x_{t, m};\theta_{t, m}) 
\end{equation*}
In the Poisson GVAR(1) process, $p$ is the Poisson probability mass function with a canonical parameter $\theta_{t, m}$ that is a linear combination of the lagged values $X_{t - 1, m}$: 
\begin{equation*}
p(x_{t, m}; \theta_{t, m}) = \frac{\exp\{x_{t, m} \theta_{t, m} - \exp(\theta_{t, m})\}}{x_{t, m}!}
\quad\text{and}\quad
\theta_{t, m} = \nu_m + a_{m}' x_{t - 1} 
\end{equation*}
We denote by $A$ the $M\times M$ matrix whose rows are the coefficients $a_m'$, and by $\nu$ the $M\times 1$ vector of constants. The vector $\nu$ is an intercept term that characterizes the (log) mean of $X_t$ in the absence of influence from past values. Typically, the matrix $A$ is the parameter of interest, as it characterizes the time dependence of the process mean on past values, as shown in Fig. \ref{fig:var-process}. Each row $a_m'$ characterizes the influence of the process history on the mean of the $m$th element of $X_t$; the sign and magnitude of $a_{mj}$ indicates the direction and strength of influence of $X_{t - 1, j}$ on $X_{t, m}$. The sparsity pattern of $A$ describes exactly the conditional dependence structure of the process: if $a_{mj} = 0$ then $X_{t, m}$ is conditionally independent of $X_{t - 1, j}$ given the other elements $X_{t - 1, \neg j}$.

\begin{figure}[ht]
    \centering
    \input{figures/fig-gvar-graph}
    \color{black}
    \caption{Graphical illustration of a GVAR(1) process and the corresponding parameter matrix $A$. In the directed graph, edges are drawn when the value of one element in the past influences the mean of another element in the present, which occurs exactly when the corresponding entry of $A$ is nonzero.}
    \label{fig:var-process}
\end{figure}

A Poisson GVAR(1) process is identifiable for any values of the parameters $\nu$ and $A$ provided $T > 1$ but moments do not exist without suitable constraints on $A$ \citep{ruiz2023graphical}. The conditional means are log-linear in past values:
\begin{equation*}
\ln\left\{\mathbb{E}(X_t\;|\;X_s, s<t)\right\} = \nu + A X_{t - 1}
\end{equation*}
The log-likelihood $\ell(\nu, A; x)$ of an observed time series $x$ can be written directly from the process definition above. 

\subsection{LASSO estimation of parameter support set}

Maximum likelihood estimates can be computed using standard techniques for generalized linear models by arranging the time series into a single long pseudo-response vector, forming a block-diagonal pseudo-covariate-matrix containing the lagged values, and estimating a regression pseudo-parameter $\beta$ that comprises a column-stacked ``vectorization'' of the process parameters $\nu, A$:
\begin{equation*}
\beta = \text{vec}(\nu \; A)'
\end{equation*}
The details of this representation are provided in the Supplement. In this work we focus on estimating the support set of the parameter matrix $A$, which we express as the set of indices designating the nonzero entries in $\beta$:
\begin{equation*}
S = \left\{j: \beta_j \neq 0\right\}
\end{equation*}
If $\hat{\beta}^\lambda$ is the LASSO estimate of $\beta$ with regularization hyperparameter $\lambda$, the corresponding LASSO estimator of $S$ is:
\begin{equation*}
\hat{S}^\lambda = \{j: \hat{\beta}_j^\lambda \neq 0\}
\quad\text{where}\quad
\hat{\beta}_j^\lambda = \argmin_\beta \left\{-\ell(\beta; x) + \lambda \|\text{vec}(A)\|_1\right\}
\end{equation*}
The Supplement provides detailed explanations and algorithms for computing LASSO estimates $\hat{\beta}^\lambda$ in the GVAR setting, following the methods of \citet{friedman2007pathwise, friedman2010regularization}. Here we note that care must be taken when applying regularization constraints so as not to penalize the intercept terms distributed throughout the pseudo-parameter vector (unlike in a standard regression problem) and the sparsity of the pseudo-covariate-matrix should be leveraged for computational efficiency. 

The standard approach to sparse estimation using the LASSO is to select the regularization hyperparameter $\lambda$ by optimizing a criterion such as goodness of fit or forecast accuracy, often through cross-validation. To simplify presentation of this approach and our subsequent proposals, we introduce the shorthand notation in Table \ref{tbl:ops-notation} for key operations.

\begin{table}[H]
\singlespacing\centering
\caption{Notations used to describe key operations in estimation algorithms.}
\begin{tabular}{c|p{2in}|l}
\textbf{Notation}
&\textbf{Meaning}
&\textbf{Example}
\\\hline

$\text{resample}(\;\cdot\;)$
&subsample/resample data
&$\text{resample}(\text{data}) = (\text{train}, \text{test})$
\\

$\hat{S}(\;\cdot\;; \lambda)$ 
&compute LASSO supports with penalty $\lambda$
&$\hat{S}(\text{train}; 0.012) = \{1, 3\}$
\\

$\hat{\beta}(\;\cdot\;; S)$ 
&compute MLEs of parameters in support set $S$ 
&$\hat{\beta}(\text{train}; \{1, 3\}) = (6.1 \; 0.76\; 0 \cdots\;0)^T$
\\

$e(\;\cdot\;; \beta)$ 
&compute an error metric for parameter value $\beta$ 
&$e(\text{test}; (6.1 \;0\; 0.76\; 0\; \cdots\;0)^T) = 153.8$
\\\hline

\end{tabular}
\label{tbl:ops-notation}
\end{table}

In practice an analyst might minimize an error metric $e^\lambda = e\{\text{test}\;;\hat{\beta}^\lambda(\text{train})\}$ by a grid search over values of $\lambda$, where the train/test data partitioning is employed to reduce the bias of the error estimate. If sample size is limited, the data may be repeatedly partitioned, an error $e^\lambda_j = e\{\text{test}_j\;;\hat{\beta}^\lambda(\text{train}_j)\}$ estimated for each partition $j = 1, \dots, J$,  and average error across partitions $J^{-1}\sum_j e_j^\lambda$ chosen as the quantity to optimize.

However, we note that optimizing fit and/or forecasts for the LASSO estimator does not, in general, produce a good estimator for the support set $S$, especially for vector autoregressive models \citep{arnold2007temporal, davis2016sparse, shojaie2010discovering}. Due to the dense parametrization --- the number of parameters $DM(M+1)$ scales quadratically in the process dimension $M$ and linearly in the model order $D$ --- often the model best approximating the true data-generating process is extremely sparse. If so, strong regularization is needed to accurately recover $S$, which induces heavy bias in the LASSO estimator. In this circumstance, the hyperparameter $\lambda$ corresponding to the best estimate of $S$ produces a poorly-fitting model that yields strongly biased predictions. As a consequence, the ``naive" approach to choosing $\lambda$ by cross-validation over-specifies $S$ and yields a high number of false positives (entries of $A$ that are in fact zero but estimated to be nonzero).

We therefore consider as a benchmark method a modification of the standard approach wherein the LASSO penalty is used to determine candidate supports but the selection of an optimal model is based on unpenalized maximum likelihood estimates. This approach is shown in detail as Algorithm \ref{alg:crossval}; in short, it seeks to optimize $e[\text{test};\hat{\beta}\{\text{train}; \hat{S}(\text{train}; \lambda)\}]$ by a grid search (instead of $e\{\text{test}\;;\hat{\beta}^\lambda(\text{train})\}$as described above).

\begin{algorithm}[H]
\caption{Benchmark method: selection using cross-validation.}
\label{alg:crossval}
\begin{algorithmic}\singlespacing
\REQUIRE{
\STATE data
\STATE regularization path $\Lambda$
}
\FOR{$j = 1$ to $J$}
\STATE \textit{form data partitions:} $(\text{train}_j,     
\text{test}_j) 
\leftarrow 
\text{resample}(\text{data})$ 
\FOR{$\lambda \in \Lambda$}
\STATE \textit{compute test set error:} $e_j^\lambda \leftarrow 
e\left(\text{test}_j, 
\hat{\beta}\left(\text{train}_j,
\hat{S}(\text{train}_j, 
\lambda)
\right)
\right)$ 
\ENDFOR
\ENDFOR
\STATE \textit{select optimal penalty}: $\lambda^* \leftarrow \argmin_\lambda \{J^{-1}\sum_j e_j^\lambda \}$ 
\ENSURE{parameter estimate 
$\hat{\beta}\left(\text{data}; \hat{S}(\text{data}; \lambda^*)\right)$
}
\end{algorithmic}  
\end{algorithm}

\color{black}\subsection{Support and model aggregation}\color{black}

We propose two innovations to improve upon selection using cross-validation (Algorithm \ref{alg:crossval}). First we implement the `bootstrapped LASSO' \citep{bach2008bolasso} which replaces LASSO support estimates $\hat{S}^\lambda$ with an intersection of many estimates $\hat{S}^\lambda_1, \dots, \hat{S}^\lambda_J$ calculated from subsamples of input data. In other words, replace $\hat{S}(\cdot; \lambda)$ by:
\begin{equation*}
\hat{S}_\text{agg}(\;\cdot\; ; \lambda) = \hat{S}(\text{resample}(\cdot); \lambda) \cap \cdots \cap \hat{S}(\text{resample}(\cdot); \lambda)
\end{equation*}
In contrast to the original method of \citet{bach2008bolasso}, however, we relax the strict intersection and instead introduce a threshold on the proportion of sets in the collection that contain a given index. 

Next, we propose selecting an optimal support set $\hat{S}_j^*$ on each of $J$ data partitions and aggregating the collection by returning those indices that occur in at least $100\times\gamma$\% of optimal supports. That is, if $\{\hat{S}_j^*\}$ are the optimal supports identified on each of the data partitions, then we take as a final estimate of $S$:
\begin{equation*}
\hat{S}^* = \left\{i: J^{-1}\sum_j \mathbf{1}\{i \in \hat{S}^*_j\} > \gamma\right\}
\end{equation*}
The model aggregation method is depicted in detail by the schematic in Fig. \ref{fig:model-agg}.

\begin{figure}
    \centering
    \input{figures/fig-method-diagram.tex}
    \caption{Schematic diagram of the model aggregation method. The functions $\hat{S}(\;\cdot\;;\;\cdot\;)$, $\hat{\beta}(\;\cdot\;;\;\cdot\;)$, and $e(\;\cdot\;;\;\cdot\;)$ are written with the order of arguments reversed to better indicate that the arrows mean passing the output of one function to the first argument of the next, \textit{e.g.}, $\hat{S}(\lambda_1;x^*_{11}) \rightarrow \hat{\beta}(\cdot\;;x^*_{11})$ means $\hat{\beta}(\hat{S}(\lambda_1;x^*_{11}); x^*_{11})$.}
    \label{fig:model-agg}
\end{figure}

We refer to the first modification as `support aggregation' and the second as `model aggregation'. Both are, essentially, thresholded intersection operations -- applied once to generate candidate models at the initial support estimation step and again to combine the best candidates in place of selecting a single model. Our proposed method is the combination of these two techniques and is detailed in Algorithm \ref{alg:model-agg}.

\begin{algorithm}[H]
\caption{Proposed method: selection by support and model aggregation}
\label{alg:model-agg}
\begin{algorithmic}\singlespacing
\REQUIRE{
\STATE data
\STATE regularization path $\Lambda$
\STATE tuning parameter $\gamma$
}
\FOR{$j = 1$ to $J$}
\STATE \textit{form data partitions:} $(\text{train}_j,     
\text{test}_j) 
\leftarrow 
\text{resample}(\text{data})$
\FOR{$\lambda\in\Lambda$}
\STATE \textit{compute test set error:} $e_j^\lambda \leftarrow 
e\left(\text{test}_j, 
\hat{\beta}\left(\text{train}_j,
\hat{S}_\text{agg}(\text{train}_j, 
\lambda)
\right)
\right)$
\STATE \textit{select optimal penalty:} $\lambda^*_j \leftarrow
\argmin_\lambda e_j^\lambda$
\STATE \textit{estimate support set:} $\hat{S}_j^* \leftarrow         
\hat{S}_\text{agg}(\text{train}_j, 
\lambda^*_j)$ 
\ENDFOR
\ENDFOR
\STATE \textit{filter by selection frequency:} $\hat{S}^* \leftarrow 
\left\{i: J^{-1}\sum_j \mathbf{1}\{i \in \hat{S}_j^*\} > \gamma\right\}$ 
\ENSURE{parameter estimate 
$\hat{\beta}\left(\text{data}, \hat{S}^*)
\right)$
}
\end{algorithmic}  
\end{algorithm}

\section{Simulations}\label{sec:simulations}

We present the results of a simulation study designed to compare method performance in the context of estimating the $M$-dimensional Poisson GVAR(1) model with mean vector:
\begin{equation*}
\theta_t = \ln\left\{\mathbb{E}(X_t|X_{t - 1} = x_{t - 1})\right\} = \nu + Ax_{t - 1}
\end{equation*}
The parameter of interest is $A \in \mathbb{R}^{M\times M}$ and it is assumed to be $s$-sparse ($\|\text{vec}(A)\|_0 = sM^2$). For the study, synthetic data of length $T$ were generated and each method was used to estimate the parameter $A$. The process dimension $M$, parametric sparsity $s$, and realization length $T$ were varied systematically to generate the distinct simulation settings. At each setting, several values of the parameter $A$ were generated, and for each distinct value, several data realizations were simulated.

Although the main objective is to compare our proposed method, the combination of support aggregation and model aggregation, with the benchmark of model selection via grid search, we also employed algorithms with each modification separately in order to assess the marginal effects of each algorithm feature. As such, four distinct estimation methods were applied to each dataset in the simulation study. 

\subsection{Study design}\label{subsec:sim-design}

The study is a factorial design generated by three levels of process dimensions, $M = 10, 15, 20$, and three levels of sparsity, $s = 0.01, 0.02, 0.05$. For each factorial combination of sparsity and dimension, $10$ different $A$ matrices were generated. Further, for each distinct parameter matrix $A$ three data realization lengths $T = 500, 1000, 2000$ were considered; $5$ process realizations of each length were simulated. In total, the number of datasets simulated is $5\times 3\times 10\times 3\times 3 = 1350$ (dataset replicates $\times$ realization lengths $\times$ parameter replicates $\times$ sparsity levels $\times$ dimensions).

\subsection{Parameter generation}\label{subsec:simulations-parameters}

Poisson GVAR(1) process parameters cannot be drawn at random due to the existence of regimes in the parameter space that are either (i) unstable (unbounded means and/or variances) or (ii) unidentifiable in practice (produce data realizations containing no information about the parameters with high probability). Therefore some care is required in generating parameter values for simulating data.

In our study, parameters are generated by first simulating a stable proposal and then computing a `recoverability' statistic to determine whether to accept proposals. Proposals are constructed by the following steps:
\begin{enumerate}
\item draw nonzero parameter magnitudes from a uniform  distribution on $(0, 1)$
\item allocate a sign to each parameter 
\item fix the positions of nonzero parameters to satisfy stability constraints
\end{enumerate}
To ensure that proposals are neither too difficult nor too easy to estimate, we compute an expected `recoverability' score based on the expression for the signal-to-noise ratio in linear regression and similar to that used by \citet{czanner2015measuring} to quantify signal-to-noise for GLMs. The statistic is based on deviance, which measures the difference between the log-likelihood $\ell(A, \nu; x)$ of a model with parameters $A, \nu$ and the log-likelihood $\ell_0$ of a saturated model with one parameter per observation. We denote deviance by $\text{dev}(A, \nu; x) = -2(\ell(A, \nu; x) - \ell_0)$. We then characterize parameter recoverability by the statistic:
\begin{equation*}
    \text{recoverability}(A, \nu; x)
    = \frac{\text{dev}(0, \nu; x) - \text{dev}(A, \nu; x)}
        {\text{dev}(A, \nu; x)}
\end{equation*}
This statistic is the relative change in deviance attributable to the transition matrix $A$. (It can also be viewed as a generalization of the $F$-test for overall significance in regression.) We approximate its expected value by simulation: repeatedly simulate realizations using the parameters $A$ and $\nu$ and compute the average relative change in deviance across realizations. Empirical experiments indicated a modest number ($n = 10$ replicates) of long ($T = 100,000$) simulated realizations produces a stable estimate with relatively low variance. For our simulation study, intercepts were fixed at a constant value for each setting and proposals for $A$ were accepted if the estimated expected recoverability was at least 0.5 but did not exceed 1.5 -- that is, if the null model has an expected increase in deviance of 50\%-150\% relative to the true data-generating model.

\subsection{Results}\label{subsec:sims-results}

\begin{figure}[ht]
    \centering
    \includegraphics[scale = 1]{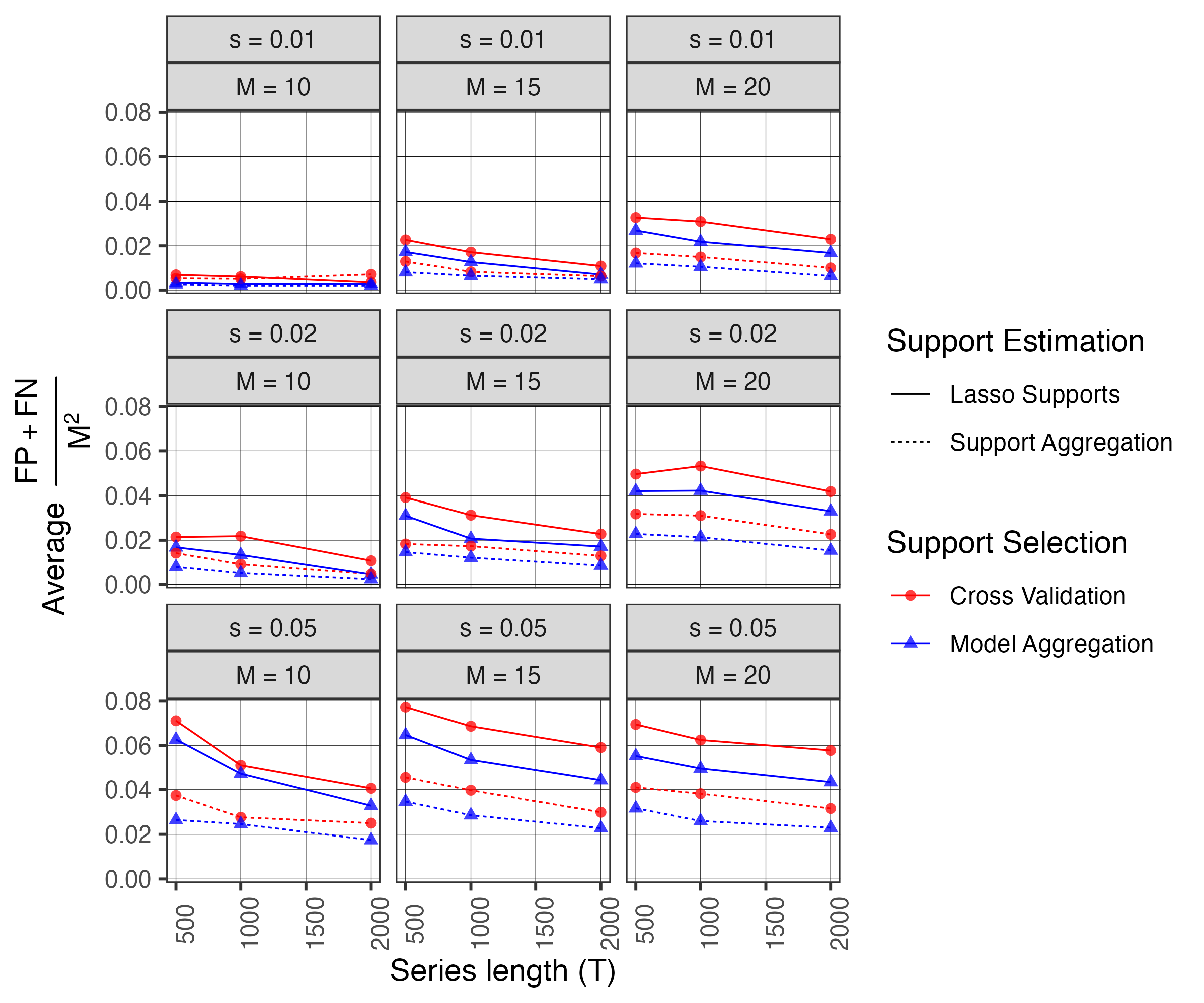}
    \caption{Average selection errors observed in simulation as measured by the proportion $\frac{FP + FN}{M^2}$ of parameters in $A$ that are incorrectly estimated as zero (FN) or nonzero (FP). Average selection errors are plotted against simulated realization length for each combination of support estimation and support selection method and each combination of sparsity $s$ (row) and process dimension $M$ (column).}
    \label{fig:selection-errors}
\end{figure}

\begin{figure}[ht]
    \centering
    \includegraphics[scale = 0.8]{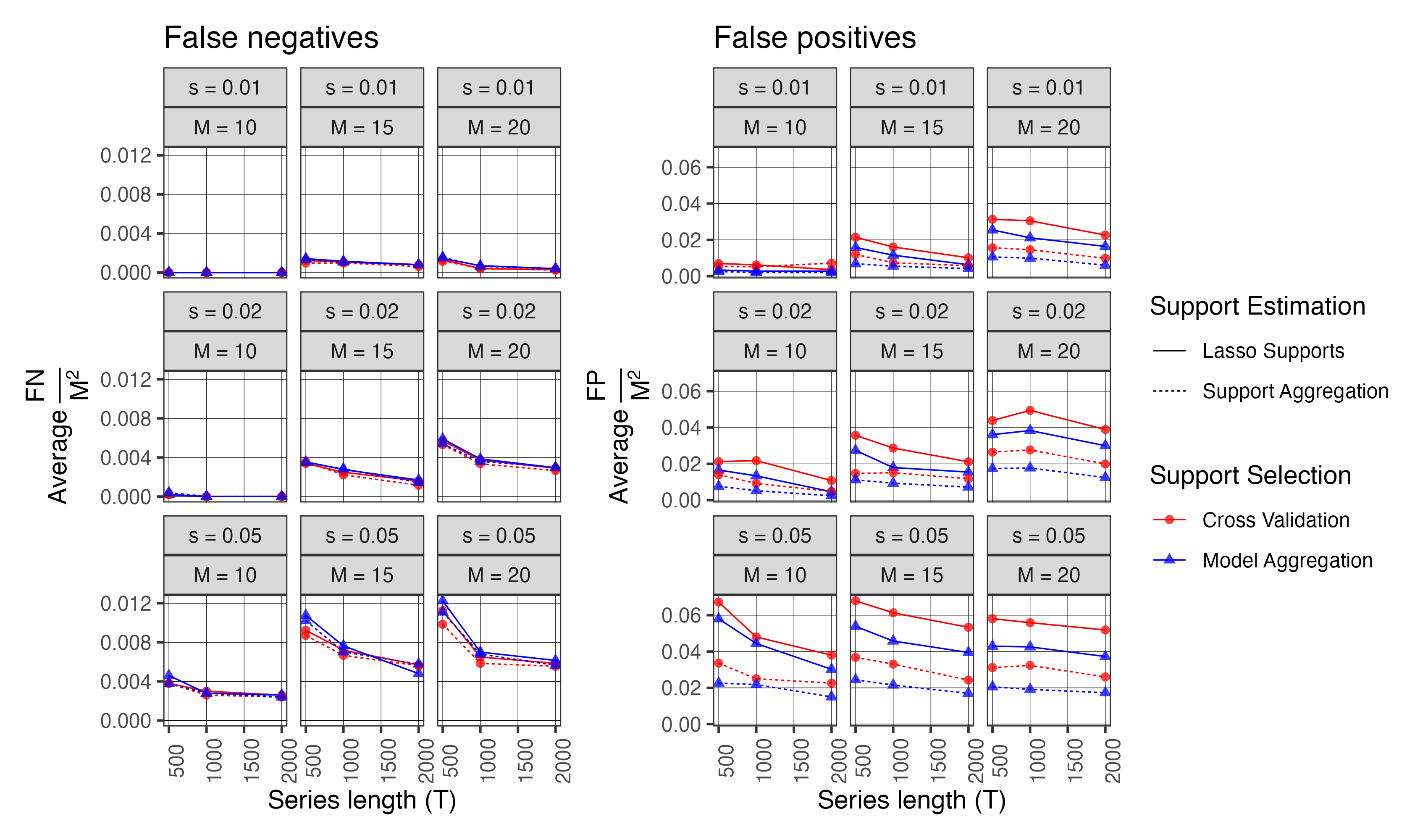}  
    \caption{Average false positive (FP) and false negative (FN) rates for each method as a function of simulated series length. Separate panels are shown for each combination of sparsity $s$ (row) and process dimension $M$ (column).}
    \label{fig:fpfn-rates}
\end{figure}

Figure \ref{fig:selection-errors} shows the average selection error for each method --- the percentage of erroneously-identified parameters in the estimated support set. Average selection error is expressed as the sum of false positives (FP) and false negatives (FN) as a proportion of the total number $M^2$ of parameters in the parameter matrix $A$. Average error rates are plotted against realization length $T$ for each combination of parameter sparsity $s$ and process dimension $M$. The lowest-dimensional and sparsest case ($M = 10$ and $s = 0.01$) represents selection and estimation of a single nonzero autoregressive parameter out of a possible $100$; and the highest-dimensional and densest case ($M = 20$ and $s = 0.05$) represents selection and estimation of $20$ nonzero parameters out of a possible $400$. For all methods, the average error rates range from $0$ to $10\%$ across combinations of $M, s, T$, and selection errors decrease with $T$ and increase with $M$ and $s$. Only in the lowest-dimensional and sparsest setting do all methods perform equally well on average. Otherwise, support aggregation outperforms the LASSO and model aggregation outperforms model selection; the performance gaps widen as $s$ and $M$ increase. The combination performs favorably, maintaining a $0-2\%$ error rate across cases. These improvements are driven largely by better control of false positive rates, as shown in Figure \ref{fig:fpfn-rates}.

\begin{figure}[ht]
    \centering
    \includegraphics[width=1\linewidth]{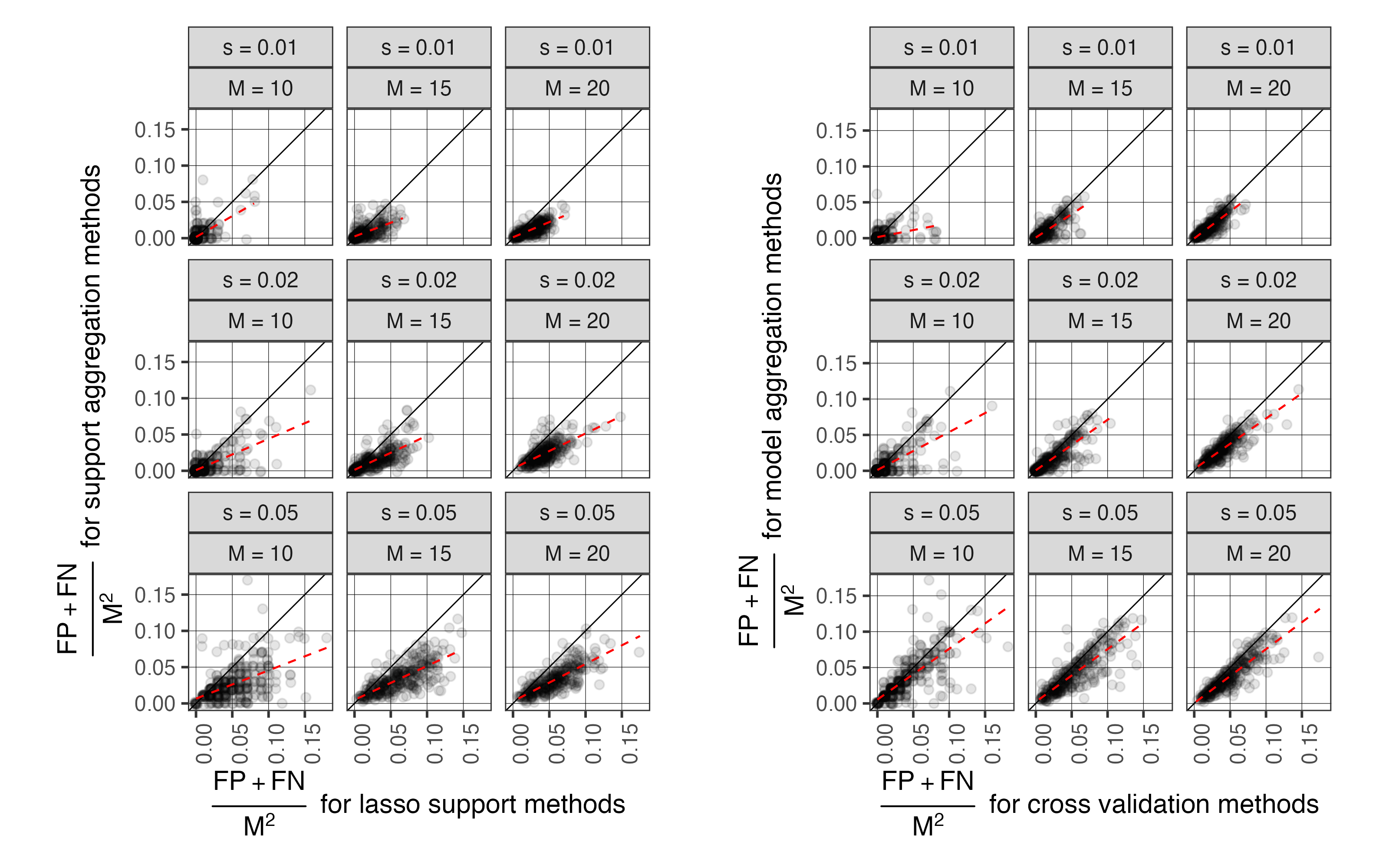} 
    \caption{Marginal effect of algorithm features on average selection error as measured by the proportion $\frac{FP + FN}{M^2}$ of parameters in $A$ that are incorrectly estimated as zero (FN) or nonzero (FP). Left: effect of support aggregation relative to LASSO estimation on average selection error. Right: effect of model aggregation relative to cross validation on average selection error. Separate panels are shown for each combination of sparsity $s$ (row) and process dimension $M$ (column).}
    \label{fig:marginal-effects}
\end{figure}

Figure \ref{fig:marginal-effects} shows the marginal effect of each methodological innovation on a dataset-by-dataset basis. At left, the error rate from the support aggregation method is plotted against the error rate from the LASSO support estimation method keeping other algorithm features fixed; these comparisons are shown for every dataset generated in the simulation study and arranged in panels according to the combination of sparsity $s$ and dimension $M$ in the same layout as Figure \ref{fig:selection-errors}. At right, the error rates for model aggregation are compared with those for cross validation in the same manner. Simple regression lines are superimposed to help visualize trends, along with a $y = x$ reference that indicates equal performance. The trends, and in fact the majority of the points themselves, are below the reference line, indicating that each innovation outperforms its benchmark independently of other algorithm features.

We note that mean square estimation error (MSE) was found not to differ appreciably between the methods (Supplemental Fig. S1). This suggests that the estimated sparsity patterns differ primarily on parameter estimates with small absolute values.

\section{Application}\label{sec:application}

\begin{figure}
    \centering
    \includegraphics[width=0.7\linewidth]{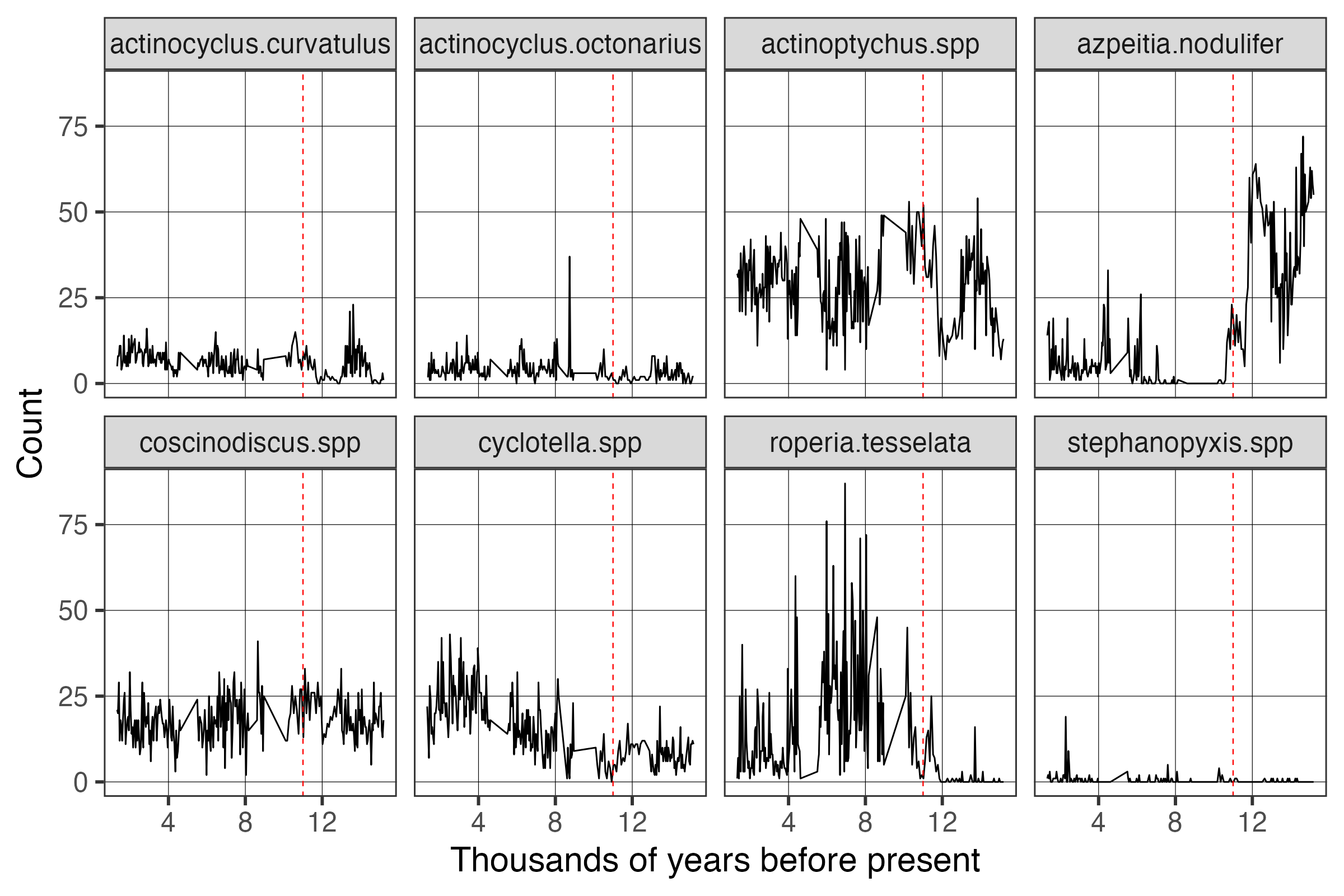}
    \caption{Counts of the numbers of individuals over time identified at the taxon level for eight diatom taxa present in marine sediment core samples from the Guayamas basin in Baja California; sea surface temperature reconstructions show a dramatic change following the end of the Pleistocene epoch, indicated by the dashed red line.}
    \label{fig:diatom-counts}
\end{figure}

We demonstrate application of our methodology using paleoclimatology data from \citet{barron2005paleoceanographic}. Data are publicly available online through the World Data Service for Paleoclimatology managed by the National Centers for Environmental Information at NOAA \cite{barron2005data}. The data are multivariate counts of diatoms from eight taxa over time and are shown in Fig. \ref{fig:diatom-counts}. Data were collected from samples taken at evenly-spaced depths in marine sediment cores from the Guayamas basin in Baja California; the depths sampled yield approximately homogeneous time intervals of 40-60 years between observations, and in all, the data span 16,000 years. 

The observation window covers the transition from the Pleistocene epoch to the Holocene epoch around 11,700 before present. Sea surface temperature reconstructions from the sampling location reflect a shift toward warmer average temperatures occurring at approximately that time, and Fig. \ref{fig:diatom-counts} reflects a changepoint, especially visible for \textit{roperia tesselata} and \textit{azpeitia nodulifer}. We speculated that this shift suggested a nonstationarity in the underlying process, and applied our methods to fit a piecewise model by epoch to investigate whether the dependence structure would reflect a corresponding change. We hypothesized that if a change in dependence structure were indeed present, the improved selection performance of our method would make such a shift more visible than benchmarks.

We fit piecewise Poisson GVAR(1) models to the data by epoch to estimate interaction networks describing serial interdependence between diatom taxa. We did not consider higher-order dependence due to the sparsity of the order-one model across estimation methods. By partitioning the data according to geological epoch and fitting models piecewise to each partition, we identified a clear and apparent change in the connectivity structure before and after the transition between geological epochs using our method. This same change was not identifiable using alternative estimation methods.

We compared three estimation methods: a naive direct application of the LASSO as described in the Methods section; the benchmark method from our simulation study and described in Algorithm \ref{alg:crossval}; and the aggregation method described in Algorithm \ref{alg:model-agg}. Results are shown in Fig. \ref{fig:diatom-graphs}, and parameter estimates are reported in Supplemental Table S1 and Supplemental Table S2. Due to its sparsity, only the aggregation method clearly supports the conclusion that there is a change in connectivity structure. Although there are some differences in connectivity as estimated by the alternative methods, there also remain inter-taxon links that persist across epochs that introduce ambiguity in interpreting the results; there is no such ambiguity using our method.

\begin{figure}
    \centering
    \includegraphics[width=1\linewidth]{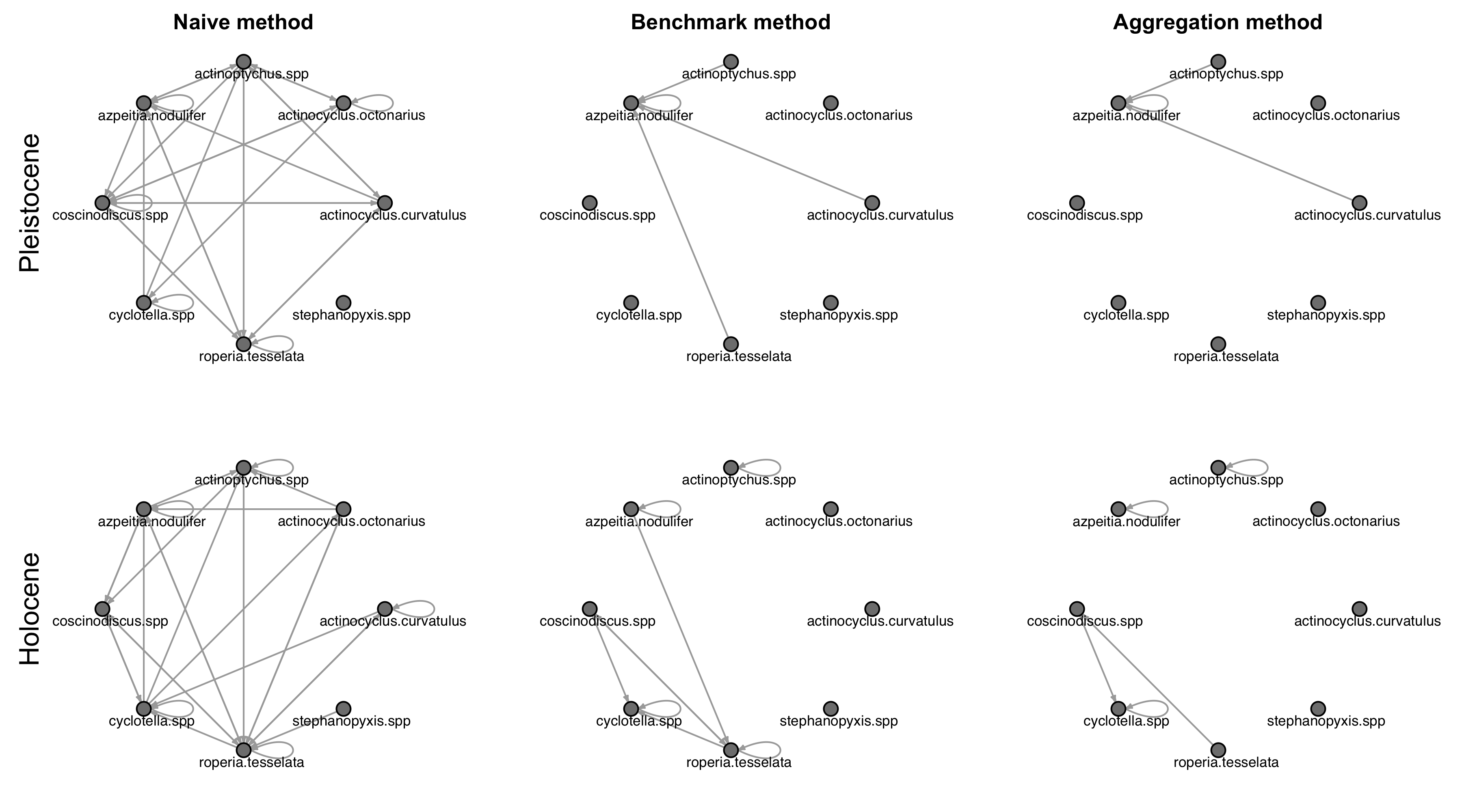}
    \caption{Interaction graphs representing inter-taxon dependencies before and after the climate change event as estimated by three methods: a `naive' method that consists in directly applying the LASSO at left; the benchmark considered in our simulation study in center; and our proposed method at right. Only the proposed method identifies a clear and apparent change in connectivity.}
    \label{fig:diatom-graphs}
\end{figure}

\section{Discussion}\label{sec:discussion}

\subsection{Summary}

This paper presents novel methodology for estimation of parameter supports using resampling and aggregation operations (Algorithm \ref{alg:model-agg}; Fig. \ref{fig:model-agg}
). The methodology is fully general but developed in the context of Poisson GVAR models for multivariate count-valued time series (Fig. \ref{fig:var-process}). We provide empirical validation demonstrating a consistent improvement over benchmark methods across process and parameter dimension, sparsity level, and series length or sample size; not only is there a consistent average effect (Fig. \ref{fig:selection-errors}, Fig. \ref{fig:fpfn-rates}), but the method outperforms alternatives for the vast majority of simulated datasets (Fig. \ref{fig:marginal-effects}). The simulation study reveals that most of the advantage of our method lies in better control of false positive rates in estimating parameter supports. We illustrate the practical advantages of this behavior with a data analysis involving estimating ecological interaction networks and comparing the connectivity structures for differences. While there is no ground truth in this context, the greater sparsity of estimates resulting from our method makes apparent differences in network structure that are otherwise obscured by additional parameters that reflect either errors or weak correlations.

It is worth highlighting that all of the benchmark methods we consider in our simulation study involve using maximum likelihood refits of LASSO estimates rather than the LASSO estimates themselves. Although we exclude direct LASSO estimates from consideration in the simulations, we illustrate the effect of refitting in our data analysis by including the `naive' method of direct LASSO estimation in the data application; the naive method produces a nearly-fully-dense model. Anecdotally, similar behavior was observed with simulated data in the course of developing the simulation study --- the naive LASSO estimate that optimized deviance produced an unacceptably high selection error rate. Thus, the benchmarks in this paper appear to represent a considerable improvement over the methods of \cite{mark2017network, hall2018learning, pandit2020generalized} in the small-sample and low-dimensional setting. We note that the notion of ``small-sample'' is relative to the process dimension $M$ in this context. All of the results in our paper concern the regime in which the number of parameters ($M(M+1)$) does not substantially exceed the series length $T$; estimation in the $M(M + 1) \gg T$ regime, while possible with our approach, may present additional challenges.

It is an important feature of our method that support sets are re-estimated across the regularization paths for each resample, rather than fixed based on a single estimate from the full dataset or some other heuristic. This is a key difference between our approach and \citet{ruiz2020sparse}. We found that conditioning our procedure on a fixed collection of candidate supports, although dramatically less computationally burdensome, produces a noteworthy deterioration of selection performance.

\subsection{Hyperparameter specification}

The resampling and aggregation approach overcomes the problem of choosing exactly one regularization hyperparameter $\lambda$. For most applications, utilizing standard heuristics for fixing $\Lambda = [\lambda_\text{min}, \lambda_\text{max}]$ and utilizing 10-20 resamples and $\gamma \in (0.6, 0.9)$ should provide an adequate starting point. Since our methods build on the LASSO estimate, we would expect improvements independently of how carefully the set of possible values $\Lambda$ is chosen; however, the model aggregation algorithm is much more sensitive to the regularization path $\lambda_1, \dots, \lambda_k\in\Lambda$ than to the threshold hyperparameter $\gamma$. We note that $\Lambda$ need not span the null model to the fully dense model, and considering that computation time increases with the number of $\lambda$ values used, we chose to perform some exploratory fits to approximate the relationship between the penalty $\lambda$ and the number $|\hat{S}|$ of parameters selected and focus on a narrow region where $|\hat{S}|$ varies the most. However, any heuristic for specifying LASSO regularization parameters would be appropriate. We note that due to the dense parametrization of GVAR models, performance can be quite sensitive to too much spacing between $\lambda$ values spanning $\Lambda$, and care should be taken to specify an appropriate value grid. Due to the above-noted sensitivity, adjusting the regularization path is most likely to provide a means of handling unexpected results.

With regard to the number of resamples used in support and model aggregation, we found that relatively low numbers of resamples --- around 10 --- were sufficient for stable performance across settings. Our approach can exploit parallelism to maintain parity with benchmark methods in computation time. Computation times for our implementation observed in the simulations are shown in Supplemental Fig. S2. In general, computation time increased exponentially in series length $T$; for very large datasets, our method could mitigate this burden by selecting resamples of substantially shorter length than the full series. No such option is available for the cross-validation approach.

We cross-validated the entire estimation procedure(s) to select an optimal threshold parameter $\gamma$ for model aggregation. This outer validation step has relatively little impact on computation time, and as a result the limitation on candidate values considered is really the number of resamples. In our case, using 10 resamples, the values giving unique solutions were $\gamma = 0.1, 0.2, \dots, 0.9, 1$; we considered all those exceeding 0.5, \textit{i.e.}, all estimates in which parameters selected more often than not are included in the support set.

\subsection{Time series resampling}

Although it is under-emphasized throughout the paper, the resampling of multiple time series can be accomplished in a number of ways. Ideally, if enough data are available, non-overlapping subsamples could be used to minimize the correlation between estimates on each subsample due to long-range dependence in the data. In our implementations, we subdivided data evenly into blocks and omitted one block at a time, concatenating the remaining data and introducing one discontinuity point across which the subsampled process is no longer stationary; the concatenated data was used for training and the held-out block was used for testing. Thus, for our method, increasing the number of resamples taken diminishes the length of series used for validation. One might instead use a block bootstrap or a residual bootstrap method such as described by \citet{kreiss2012bootstrap}.

\subsection{Limitations}

The simulation of stable process realizations without artificial thresholds proved to be challenging, as there is no obvious necessary and sufficient condition on the parameters that ensures moments exist. Furthermore, some process parametrizations are extremely challenging to recover from data because the likelihood varies little across large neighborhoods with high probability. In the course of developing empirical validation of our methodology, we therefore also developed heuristics for selecting process parameters that are both stable and recoverable. However, in using these heuristics we excluded regions of the parameter space from consideration in our simulations. We also note that our simulation results do not account for the possibility of model misspecification. Although our method can be extended directly to estimate higher-order processes by leveraging the fact that any order-$D$ process has an order-1 representation, we did not explore the problem of order selection in this work, and assumed throughout that the order of autoregression is correctly specified. In theory, the model aggregation approach could be extended to handle misspecification by replacing the LASSO with an alternative method that also performs order selection, or by introducing groupwise penalties on parameters associated with different lags. Further work could explore this topic.

\newpage
\bibliographystyle{plainnat}
\bibliography{refs}

\newpage
\spacing{1}
\renewcommand{\thefigure}{S\arabic{figure}}
\renewcommand{\thetable}{S\arabic{table}}
\renewcommand{\thealgorithm}{S\arabic{algorithm}} 
\setcounter{algorithm}{0}
\setcounter{figure}{0}
\setcounter{table}{0}

\begin{appendices}
\section{Algorithms}

The algorithms used to produce estimates for numerical experiments and data analysis in the main body of the paper are presented in detail in this appendix. Essentially, we apply the coordinate descent method of \citet{friedman2010regularization} after manipulating the likelihood optimization problem into a form similar to univariate regression and partitioning coordinates into blocks for computational efficiency. The core technique is based on coordinate descent updates for the weighted least squares regression estimate with an elastic net penalty. We first present this method, with full derivations, then the approach of \citet{friedman2010regularization}, and lastly the details of our implementation for GVAR(D) estimation.

\parheader{Weighted least squares with elastic net} Consider estimation of the regression model
\begin{equation*}
    z = X\beta + \epsilon
\end{equation*}
where $z$ is the response, $X$ is the matrix of fixed covariate information, $\beta$ is a vector of fixed but unknown parameters, and $\epsilon$ is a vector of uncorrelated random errors with mean zero. 

The weighted least squares estimate of $\beta$ is given by 
\begin{equation*}
\hat{\beta} = \argmin_\beta \left\{
		\frac{1}{2}(z - X\beta)' W (z - X\beta)
	\right\}
\end{equation*}
where $W$ is a diagonal matrix of known weights. Denote the objective function by 
\begin{equation}\label{eqn:app_alg_wlsobjective}
f(\beta) = \frac{1}{2}(z - X\beta)' W (z - X\beta)
\end{equation}
The weighted least squares estimate with an elastic net penalty is obtained by adding a penalty term $P(\beta)$ to the objective function 
\begin{equation}
\label{eqn:app_alg_enetwlsobjective}
\hat{\beta} = \argmin_\beta \left\{
		\frac{1}{2}(z - X\beta)' W (z - X\beta) 
			+ \lambda P(\beta)
	\right\}
\end{equation}
where $P(\beta) = \frac{1 - \alpha}{2}\|\beta\|_2^2 + \alpha\|\beta\|_1$ and $\lambda$ is a hyperparameter that controls the strength of the penalty. The hyperparameter $\alpha$ controls the relative balance of the $L_2$ (ridge) and $L_1$ (LASSO) terms in the penalty.

The coordinate descent approach to computing $\hat{\beta}$ in Equation \eqref{eqn:app_alg_enetwlsobjective} is to iteratively minimize the objective function in each $\beta_j$ conditional on the current values of the other parameters $\beta_k$ for $k \neq j$ until convergence. Each iteration can be accomplished by a simple update rule that can be equivalently derived by a number of approaches \citep{zou2005regularization, van2007prediction, friedman2007pathwise, friedman2010regularization}. Here the rule is derived from the update given by Newton's method.

For the purpose of taking derivatives in $\beta_j$, the first portion of the objective function $f(\beta)$ can be rewritten as
\begin{equation*}
f(\beta)
= \frac{1}{2}\left(z - \sum_{k \neq j} x_k\beta_k - x_j\beta_j\right)'
	W \left(z - \sum_{k \neq j} x_k\beta_k - x_j\beta_j\right)
\end{equation*}
where $x_j$ denotes the $j$th column of $X$. The first and second partial derivatives of this portion with respect to $\beta_j$ are
\begin{align*}
\frac{\partial}{\partial\beta_j}f(\beta)
    &= - \left(z - \sum_{k \neq j} x_k\beta_k - x_j\beta_j\right)' W x_j \\
    &= \beta_j x_j' W x_j - \left(z - \sum_{k \neq j} x_k\beta_k\right)' W x_j\\
    &= \beta_j x_j' W x_j - \left(z - z^{(j)}\right)' W x_j \\
\frac{\partial^2}{\partial^2\beta_j}f(\beta)
    &= x_j' W x_j
\end{align*}
where $z^{(j)}$ denotes $\sum_{k \neq j} x_k\beta_k$. The quantity $\left(z - z^{(j)}\right)$ is known as the $j$th partial residual.

Now, the derivatives of the penalty portion of the objective function are
\begin{align*}
    \frac{\partial}{\partial\beta_j}\left(\lambda\frac{1 - \alpha}{2}\|\beta\|_2^2 + \lambda\alpha\|\beta\|_1\right)
        &= \lambda(1 - \alpha)\beta_j + \lambda\alpha \frac{\partial}{\partial\beta_j}|\beta_j| \\
        &= \begin{cases}
            \lambda(1 - \alpha)\beta_j + \lambda\alpha &\text{ , if } \beta_j > 0 \\
            \lambda(1 - \alpha)\beta_j - \lambda\alpha &\text{ , if } \beta_j < 0 \\
            \lambda(1 - \alpha)\beta_j + \lambda\alpha\partial|\beta_j| &\text{ , if } \beta_j = 0
        \end{cases} \\
    \frac{\partial^2}{\partial^2\beta_j}\left(\lambda\frac{1 - \alpha}{2}\|\beta\|_2^2 + \lambda\alpha\|\beta\|_1\right)
        &= \lambda(1 - \alpha) 
\end{align*}
where $\partial|\beta_j|$ denotes a subgradient. Supposing the current estimate is $\hat{\beta}^{(k)}$, the Newton update for minimization in the $j$th coordinate is
\begin{equation}\label{eqn:app_alg_newtonupdate}
    \hat{\beta}^{(k + 1)}_j
    \longleftarrow
    \hat{\beta}^{(k)}_j 
        - \frac{\frac{\partial}{\partial\beta_j}\left(f(\beta) + \lambda P(\beta)\right)\rvert_{\beta = \hat{\beta}^{(k)}}}
        {\frac{\partial^2}{\partial^2\beta_j}\left(f(\beta) + \lambda P(\beta)\right)\rvert_{\beta = \hat{\beta}^{(k)}}}
\end{equation}
When articulated casewise according to whether $\hat{\beta}^{(k)}_j > 0$ or $\hat{\beta}^{(k)}_j < 0$ or $\hat{\beta}^{(k)}_j = 0$, Equation \eqref{eqn:app_alg_newtonupdate} gives the update rule (after minor algebraic simplification)
\begin{equation*}
    \hat{\beta}_j^{(k + 1)} 
    \longleftarrow
    \begin{cases}
        \frac{\left(z - z^{(j)}\right)' W x_j - \lambda\alpha}{x_j' W x_j + \lambda(1 - \alpha)} &\text{ , if } \hat{\beta}_j^{(k)} > 0 \\
        \frac{\left(z - z^{(j)}\right)' W x_j + \lambda\alpha}{x_j' W x_j + \lambda(1 - \alpha)} &\text{ , if } \hat{\beta}_j^{(k)} < 0 \\                \frac{\left(z - z^{(j)}\right)' W x_j - \lambda\alpha c}{x_j' W x_j + \lambda(1 - \alpha)} 
            &\text{ , if } \hat{\beta}_j^{(k)} = 0 \\
    \end{cases}
\end{equation*}
where $c \in [-1, 1]$. Now if $|\left(z - z^{(j)}\right)' W x_j| \leq \lambda\alpha$, the (sub)gradient in Equation \eqref{eqn:app_alg_newtonupdate} is not a descent direction, and in this case, the optimal value of the objective function is achieved by setting $\hat{\beta}_j^{(k + 1)}$ to zero. Rewriting the conditions in terms of $\left(z - z^{(j)}\right)' W x_j$ and dropping the iteration count $k$ yields the update rule
\begin{equation*}
    \hat{\beta}_j
    \longleftarrow
    \begin{cases}
        \frac{\left(z - z^{(j)}\right)' W x_j - \lambda\alpha}{x_j' W x_j + \lambda(1 - \alpha)} 
            &\text{ , if } \left(z - z^{(j)}\right)' W x_j > 0 \text{ and } |\left(z - z^{(j)}\right)' W x_j| > \lambda\alpha \\
        \frac{\left(z - z^{(j)}\right)' W x_j + \lambda\alpha}{x_j' W x_j + \lambda(1 - \alpha)} 
            &\text{ , if } \left(z - z^{(j)}\right)' W x_j < 0 \text{ and } |\left(z - z^{(j)}\right)' W x_j| > \lambda\alpha \\
        0 &\text{ , if } \text{ and } |\left(z - z^{(j)}\right)' W x_j| \leq \lambda\alpha 
    \end{cases}
\end{equation*}
Finally, the update can be re-expressed in terms of the soft-threshold function
\begin{equation*}
    S(x, y) 
    = \text{sign}(x)(|x| - y)_+
    = \begin{cases}
        x - y &\text{ , if } x > 0 \text{ and } |x| > y \\
        x + y &\text{ , if } x < 0 \text{ and } |x| > y \\
        0 &\text{ , if } |x| \leq y 
    \end{cases}
\end{equation*}
compactly as the coordinate update
\begin{equation*}
\hat{\beta}_j
\longleftarrow
\frac{S\left(
    \left(z - z^{(j)}\right)' W x_j,
    \lambda\alpha\right)}
	{x_j' W x_j + \lambda(1 - \alpha)}
\end{equation*}
The coordinate descent algorithm using this update rule and a gradient norm stopping criterion is given as Algorithm \ref{alg:app_enetwls}.
\begin{algorithm}
\caption{Coordinate descent algorithm for computing the weighted least squares estimate with an elastic net penalty.}
\label{alg:app_enetwls}
\begin{algorithmic}
\REQUIRE{
\STATE Data $z, X$; hyperparameters $\lambda, \alpha$; and convergence threshold $\delta$
}
\STATE Initialize $\hat{\beta}$
\WHILE{$\|\nabla f(\hat{\beta})\|_2 > \delta$}{
\FOR{$j = 1, \dots, J$}
    \STATE $z^{(j)} \longleftarrow \sum_{k \neq j} x_k\hat{\beta}_k$
    \STATE  $\hat{\beta}_j 
        \longleftarrow
        \frac{S\left(\left(z - z^{(j)}\right)' W x_j, \lambda\alpha\right)}
        {x_j' W x_j + \lambda(1 - \alpha)}$
\ENDFOR}
\ENDWHILE
\ENSURE $\hat{\beta}$
\end{algorithmic}
\end{algorithm}

\parheader{Penalized IRLS for univariate generalized linear models} Consider the univariate generalized linear model: a response vector $y$ and covariate information $X$ where
\begin{equation*}
y_i \stackrel{indep.}{\sim} p(\cdot\; ; \theta_i)
\quad\text{and}\quad
g(\mathbb{E}y) = g(\mu) = \eta = X\beta
\end{equation*}
where $p$ is an exponential family density of the form 
\begin{equation*}
p(y_i;\theta_i) \propto \exp\left\{\frac{y_i\theta_i - b(\theta_i)}{a(\phi_i)}\right\}
\end{equation*}
Assume that $g$ is the canonical link function that results from equating $\eta = \theta$. The log-likelihood as a function of the linear predictor $\eta$ is, up to a constant,
\begin{equation*}
\ell(\eta; x, y) = \sum_{i = 1}^n \left( \frac{y_i \eta_i - b(\eta_i)}{a(\phi_i)} \right)
\end{equation*}
The classical technique for unpenalized maximum likelihood estimation maximizes $\ell(\eta; x, y)$ iteratively by:
\begin{enumerate}
\item computing an approximation $\tilde{\ell}$ to the log-likelihood from current estimates;
\item maximizing the approximation $\tilde{\ell}$ to get new estimates;
\item repeating 1 - 2 until the estimates converge.
\end{enumerate}
Conveniently, the approximation $-\tilde{\ell}$ takes the form of the loss function for a weighted least squares problem for which a closed form solution is available, which eases calculations in step 2.

Consider now computing the penalized MLE
\begin{equation}\label{eqn:app_alg_pglmobjective}
\hat{\beta} = \argmin_\beta \left\{-\ell(\eta; x, y) + \lambda P(\beta)\right\}
\end{equation}
The same strategy for unpenalized estimation is applicable with the penalty, though generally iterative methods will replace closed form solutions for the optimization step (step 2). The objective function $-\ell(\eta; x, y) + \lambda P(\beta)$ can be approximated by the loss function for a penalized weighted least squares problem. For the elastic net penalty, the approximation can be maximized with respect to $\beta$ via Algorithm \ref{alg:app_enetwls}. The penalized MLE in Equation \eqref{eqn:app_alg_pglmobjective} with the elastic net penalty can therefore be found iteratively by:
\begin{enumerate}
\item computing the approximation $-\tilde{\ell}(\eta; x, y) + \lambda P(\beta)$ to the (negative) penalized log-likelihood from current estimates;
\item  maximizing the approximation $-\tilde{\ell}(\eta; x, y) + \lambda P(\beta)$ with respect to one coordinate via the update rule in Algorithm \ref{alg:app_enetwls} to get new estimates;
\item repeating 1 - 2 for each coordinate and cycling through the coordinates until the estimates converge.
\end{enumerate}
This procedure is given as Algorithm \ref{alg:app_alg_pirls}.

\begin{algorithm}
\caption{Coordinatewise IRLS algorithm for estimation of generalized linear models with an elastic net penalty.}
\label{alg:app_alg_pirls}
\begin{algorithmic}
\REQUIRE{data $x, y$; hyperparameters $\lambda,\alpha$; convergence threshhold $\delta$}
\STATE Initialize $\hat{\beta}$
\WHILE{$\|\nabla\ell(\hat{\eta}; x, y)\|_2 > \delta$}{
\FOR{$j = 1, \dots, J$}
    \STATE $W \longleftarrow \left(\text{diag}(g'(\hat{\mu})\right)^{-1}$
    \STATE $z \longleftarrow x\hat{\beta} - (y - \hat{\mu})W^{-1}$
    \STATE $z^{(j)} \longleftarrow \sum_{k \neq j} x_k\hat{\beta}_k$
    \STATE  $\hat{\beta}_j 
        \longleftarrow
        \frac{S\left(\left(z - z^{(j)}\right)' W x_j, \lambda\alpha\right)}
        {x_j' W x_j + \lambda(1 - \alpha)}$
\ENDFOR}
\ENDWHILE
\ENSURE $\hat{\beta}$
\end{algorithmic}
\end{algorithm}

The update rules for $z$ and $W$ in Algorithm \ref{alg:app_alg_pirls} are derived from a Taylor expansion around current estimates. Let $\hat{\eta}$ denote the linear predictor from the current estimates $\hat{\beta}$, $\ell(\eta)$ denote the log-likelihood (suppressing the data arguments), $v$ and $W$ denote the gradient and Hessian of $\ell(\eta)$. A second-order Taylor expansion of the log-likelihood about $\hat{\eta}$ gives
\begin{align*}
\ell(\eta)
&\approx \ell(\hat{\eta}) + (\eta - \hat{\eta})' \ell'(\hat{\eta}) 
	- \frac{1}{2}(\eta - \hat{\eta})' \ell''(\hat{\eta}) (\eta - \hat{\eta}) \\
&= C + (\eta - \hat{\eta})' \ell'(\hat{\eta}) 
	- \frac{1}{2}(\eta - \hat{\eta})' \ell''(\hat{\eta}) (\eta - \hat{\eta}) \\
&= C + (\eta - \hat{\eta})' v
	- \frac{1}{2}(\eta - \hat{\eta})' W (\eta - \hat{\eta}) \\
&= C^* - \frac{1}{2}(\eta - \hat{\eta} - vW^{-1})' W 
	(\eta - \hat{\eta} - vW^{-1}) \\
&= C^* - \frac{1}{2}(\eta - z)' W 
	(\eta - z)
\end{align*}
with $z = \hat{\eta} - vW^{-1}$. Now $W$ and $z$ depend on the current estimates and can be derived in general for any GLM. Since by hypothesis $\mu_i = b'(\theta_i) = g^{-1}(\eta_i)$, 
\begin{align*}
\frac{\partial\ell}{\partial\eta_i}
&= \left(\frac{y_i - b'(\theta_i)}{a_i (\phi)}\right)\frac{\partial\theta_i}{\partial\eta_i} \\
&= \left(\frac{y_i - b'(\theta_i)}{a_i (\phi)}\right)\frac{\partial\theta_i}{\partial\mu_i}\frac{\partial\mu_i}{\partial\eta_i} \\
&= \left(\frac{y_i - b'(\theta_i)}{a_i (\phi)}\right)\frac{1}{g'(\mu_i)}\frac{1}{b''(\theta_i)} \\
&= (y_i - \mu_i) \left(a_i (\phi)g'(\mu_i)b''(\theta_i)\right)^{-1}
\end{align*}
When $a_i(\phi) = 1$ and $g$ is the canonical link, $g'(\mu_i) = \frac{\partial\eta_i}{\partial\mu_i}$ and $b''(\theta_i) = \frac{\partial\mu_i}{\partial\theta_i}$, so
\begin{equation*}
\frac{\partial\ell}{\partial\eta_i} = y_i - \mu_i
\end{equation*}
and 
\begin{equation*}
\frac{\partial^2\ell}{\partial\eta_i\partial\eta_j} 
= \begin{cases}
	-\frac{1}{g'(\mu_i)} &\qquad i = j \\
	0 &\qquad i \neq j
	\end{cases}
\end{equation*}
So $W = \text{diag}(1/g'(\mu))$ and $v = (y_1 - \mu_1 \cdots y_N - \mu_N)$. Therefore, given current estimates, updating the approximation amounts to computing
\begin{align}
z &= X'\hat{\beta} - (y - \hat{\mu})W^{-1} \label{eqn:app_alg_workingresponse}\\
W &= \left(g'(\hat{\mu})\right)^{-1} \label{eqn:app_alg_weights}
\end{align}
Equations \eqref{eqn:app_alg_workingresponse} and \eqref{eqn:app_alg_weights} give the updates that appear in the outer while loop of Algorithm \ref{alg:app_alg_pirls}.

\parheader{GVAR(D) estimation} Algorithm \ref{alg:app_alg_pirls} can be modified for efficient estimation of GVAR(D) models with an elastic net penalty using a blockwise update strategy for the inner while loop.

The GVAR(D) model for data $\{ x_t \in \mathbb{R}^M \}_{t = 0}^T$ is the random process $\{ X_t \in \mathbb{R}^M \}_{t = 0}^T$ characterized by the conditional mean structure
\begin{equation*}
g\left(\mathbbm{E} \left(X_t \;|\;  X_{t - 1} = x_{t - 1}, \dots, X_{t - D} = x_{t - D}\right)\right)
= \nu + \sum_{d = 1}^D A_d x_{t - d}
\end{equation*}
under the assumption that for each $m = 1, \dots, M$ and each $t$
\begin{equation*}
X_{t, m} \;|\; X_{t - 1} = x_{t - 1}, \dots, X_{t - D} = x_{t - D} \stackrel{indep.}{\sim} p\left(\;\cdot\; \Big| \;\theta_{t, m} = \nu_m + \sum_{d = 1}^D a_{d, m}' x_{t - d} \right) 
\end{equation*}
where $p(\cdot|\theta)$ is an exponential family density and $g$ is the canonical link so that $\eta_{t, m} = \theta_{t, m} = g(\mu_{t, m}) = \mathbbm{E}(X_{t, m} | X_{t - 1}, \dots, X_{t - D})$). Denoting the data by $X = (X_0\;\cdots\; X_T)$ log-likelihood by $\ell(B; X)$, the penalized maximum likelihood estimate of $B$ is
\begin{equation*}
\hat{B} = \arg\min_{B} \left\{ 
			-\ell(B;X) + \lambda P(B)
			\right\}
\end{equation*}
where the explicit form of the likelihood depends only on the underlying density $p$. Here let $P$ be the elastic net penalty applied only to the elements of $A_1, \dots, A_D$
\begin{equation*}
P(B) = \frac{1 - \alpha}{2}\sum_{d, i, j} a_{dij}^2 + \alpha\sum_{d, i, j}|a_{dij}|
\end{equation*}
An iterative estimation algorithm based on Algorithm \ref{alg:app_alg_pirls} is derived below by vectorizing the problem and developing a blockwise update strategy.

The GVAR(D) model can be expressed as a multivariate generalized regression model 
\begin{equation*}
g\left(\mathbb{E}(Y|U)\right) = g\left(\mu\right) = \eta = UB
\end{equation*}
expressed in terms of $Y, U, B$ where
\begin{equation*}
    Y = \left[\begin{array}{c}
      X_T' \\
      X_{T - 1}' \\
      \vdots \\
      X_{D}'
    \end{array}\right]
    \quad
    U = \left[\begin{array}{cccc}
	1 &X_{T - 1}' &\cdots &X_{T - D}' \\
	1 &X_{T - 2}' &\cdots &X_{T - D - 1}' \\
	\vdots &\vdots &\ddots &\vdots \\
	1 &X_{D - 1}' &\cdots &X_{0}' \\
	\end{array}\right]
	\quad
	B = \left[\begin{array}{c}
	\nu' \\
	A_1' \\
	\vdots \\
	A_D'
	\end{array}\right]
\end{equation*}
The core model relationship can be expressed in univariate terms by applying the vectorization transformations
\begin{equation*}\label{eqn:app_alg_vectorization}
    \mathcal{Y} \stackrel{\text{def}}{=} \text{vec}Y
    \;,\quad
    \mathcal{X} \stackrel{\text{def}}{=} U \otimes I_M
    \;,\quad\text{and }
    \beta \stackrel{\text{def}}{=} \text{vec}B
\end{equation*}
with the result
\begin{equation*}
    \mathcal{Y}
    = \left[\begin{array}{c}
        x_1 \\\hdashline
        x_2 \\\hdashline
        \vdots \\\hdashline
        x_M
    \end{array}\right]
    \qquad
    \mathcal{X} 
    = \left[\begin{array}{c:c:c:c}
         U &0 &\cdots &0 \\\hdashline
         0 &U &\cdots &0 \\\hdashline
         \vdots &\vdots &\ddots &\vdots \\\hdashline
         0 &0 &\cdots &U
    \end{array}\right]
    \qquad
    \beta = \text{vec}(B)
\end{equation*}
where $x_m = (x_{0, m} \;\cdots\; x_{T, m})'$ denote the univariate series in each component so that
\begin{equation*}
    g\left(\mathbb{E}(\mathcal{Y}|\mathcal{X})\right) = g\left(\text{vec}(\mu)\right) = \text{vec}(\eta) = \mathcal{X}\beta
\end{equation*}
It is straightforward to verify by inspection that if the the elements of $\mathcal{Y}$ are assumed to be Poisson conditional on $\mathcal{X}$ with the corresponding mean, the likelihoods of $\beta$ and $B$ are identical, \textit{i.e.},
\begin{equation*}
    \ell(\beta; \mathcal{Y}, \mathcal{X}) = \ell(B; Y, U)
\end{equation*}
since the likelihood contributions from each $x_{t, m}$ are exactly the same. Therefore, the estimate $\hat{B}$ can be recovered from the estimate
\begin{equation*}
    \hat{\beta} = \argmin_\beta \left\{-\ell(\beta; \mathcal{Y}, \mathcal{X}) + \lambda P^*(\beta) \right\}
\end{equation*}
where $P^*(\beta) = P(B)$. 

Due to the sparsity pattern in $\mathcal{X}$, the computations involved in the update rule for the $l$th coordinate in the vectorized problem can be articulated as operations involving only $U$ and columns of $Y$. Specifically, denoting the row and column dimensions of $U$ as $N = T - D$ and $P = DM + 1$ and the row and column dimensions of $\mathcal{X}$ as $K = NM$ and $L = M(DM + 1)$, direct application of the calculations for the coordinate update for the $l$th position in Algorithm \ref{alg:app_alg_pirls} to the vectorized problem in terms of $\mathcal{Y}, \mathcal{X}$ gives
\begin{align*}
    \left(\mathcal{Z} - \mathcal{Z}^{(l)}\right)' \mathcal{W}\mathcal{X}_l &= \left(x_m - x_m^{(p)}\right)' W_m u_p \\
    \mathcal{X}_l' \mathcal{W} \mathcal{X}_l &= u_p' W_m u_p
\end{align*}
where $\mathcal{Z}$ is the working response vector, $\mathcal{W}$ is the diagonal matrix of update weights, and the index transformation is given by
\begin{align*}
(n, m) &= \left(k \text{mod} M + M\mathbbm{1}\{ k\text{mod}M = 0\},
				 \frac{k - k\text{mod}M}{M} + \mathbbm{1}\{k = k\text{mod}M\}
				 \right)\\
(p, m) &= \left(l \text{mod} P + P\mathbbm{1}\{ l\text{mod}P = 0\},
				 \frac{l - l\text{mod}P}{P} + \mathbbm{1}\{l = l\text{mod}P\}
				 \right)
\end{align*}
Note that the computation of $x_{mn}^{(p)} = \sum_{j \neq p} u_{nj} b_{jm}$ depends only on the $m$th column of $B$.

Consequently, a more efficient strategy is to divide the coordinate iterations into blocks corresponding to the blocks of $B$ and compute the update rules using data stored in the formats specified by $U$ and $Y$. A slight ridge penalty is included for numerical stability. This is given as Algorithm \ref{alg:app_gvar} and that which is used in the simulation experiments and data analyses presented in the main paper.

\begin{algorithm}
\caption{Coordinatewise IRLS algorithm for computing LASSO GVAR(D) estimates with blockwise updates.}
\label{alg:app_gvar}
\begin{algorithmic}
\REQUIRE{
\STATE Data $Y, U$, hyperparameters $\lambda, \alpha$, and convergence threshold $\delta$
}
\STATE Initialize $\hat{B}$
\WHILE{$\|\nabla \ell(\hat{B}; X)\|_2 > \delta$}{
\FOR{$m = 1, \dots, M$}
\FOR{$p = 1, \dots, DM + 1$}
    \STATE $W_m \longleftarrow \left(\text{diag}(g'(\hat{\mu}_m)\right)^{-1}$
    \STATE $z_m \longleftarrow U\hat{b}_m - (x_m - \hat{\mu}_m)W_m^{-1}$
    \STATE $z_{mn}^{(p)} \longleftarrow \sum_{j \neq p} u_{nj}\hat{b}_{jm}$ for $n = 1, \dots, N$
    \STATE  $\hat{b}_{pm} 
        \longleftarrow
        \begin{cases}
            \frac{\left(z_m - z_m^{(p)}\right)' W_m u_p}
            {u_p' W_m u_p} &\text{, if } p = 1 \\
            \frac{S\left(\left(z_m - z_m^{(p)}\right)' W_m u_p, \lambda\alpha\right)}
            {u_p' W_m u_p + \lambda(1 - \alpha)} &\text{, if } p > 1
        \end{cases}$
\ENDFOR
\ENDFOR}
\ENDWHILE
\ENSURE $\hat{B}$
\end{algorithmic}
\end{algorithm}

\newpage
\section{Supplemental Figures}

\begin{figure}[H]
    \centering
    \includegraphics[width=1\linewidth]{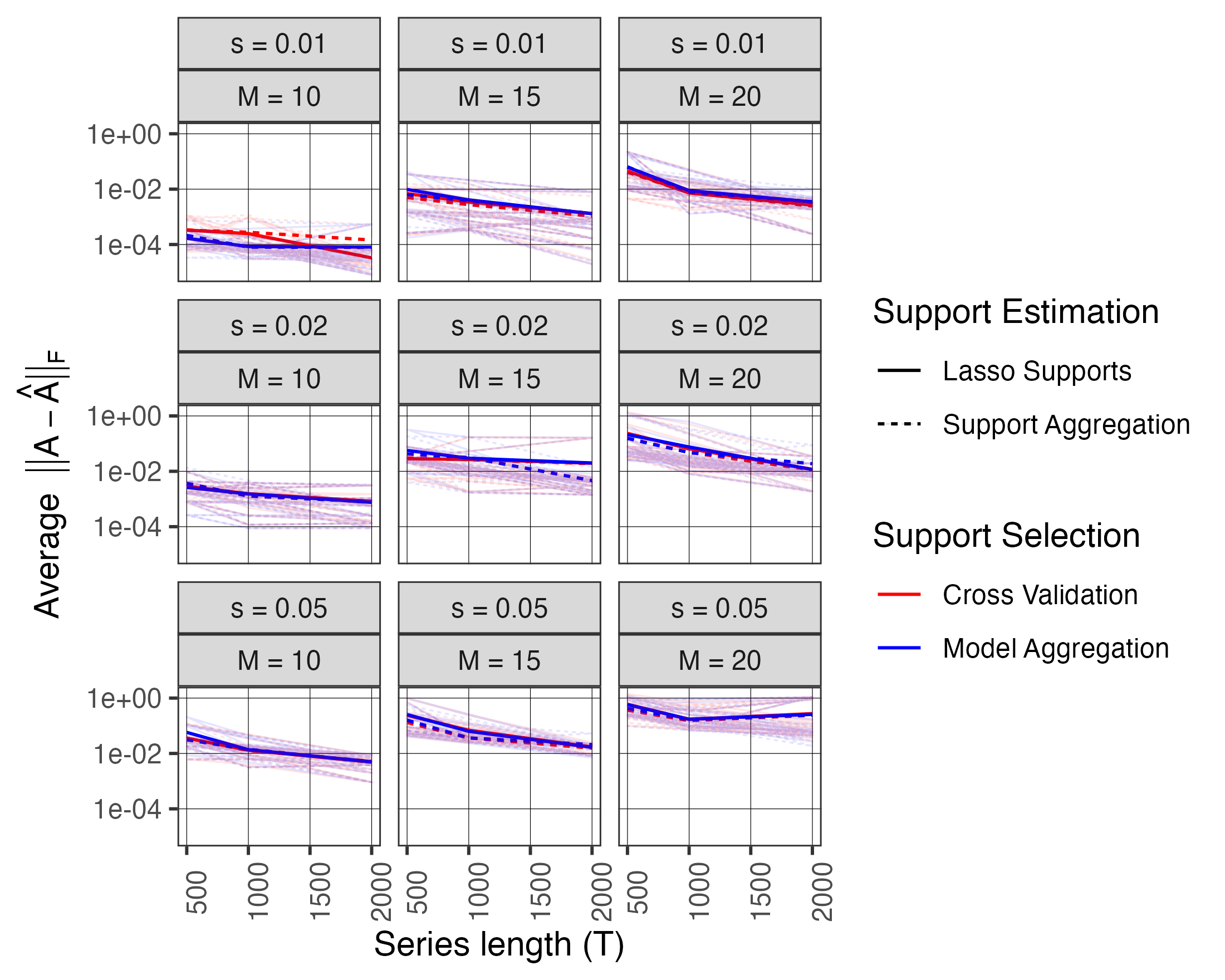}
    \caption{Mean square estimation errors (MSE) observed in the simulation described in the main portion of the paper. Faint lines indicate average MSE on a per-parameter basis (\textit{i.e.}, for each unique $A$ matrix generated) across the $5$ dataset replicates generated; bold lines show averages by simulation setting, method, and series length $T$.}
    \label{sfig:fig-mse}
\end{figure}

\begin{figure}[H]
    \centering
    \includegraphics[width=1\linewidth]{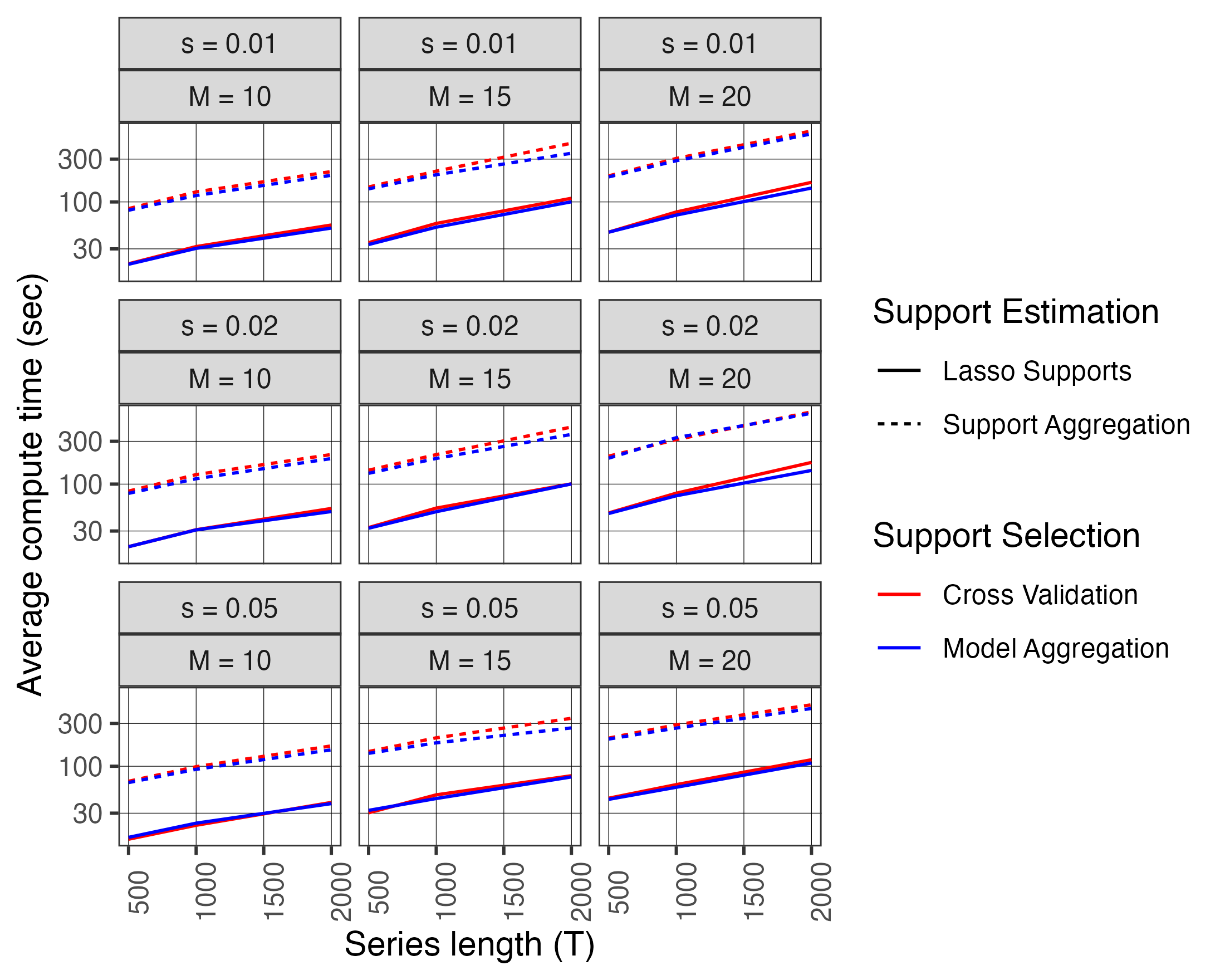}
    \caption{Average computation time in seconds, as a function of series length $T$ and shown on the $\log_{10}$ scale, to compute estimates of $A$. Averages are computed by method across the $10$ parameter matrices and $5$ datasets generated (so $n = 50$ in each average) for each simulation setting described in the main portion of the paper.}
    \label{sfig:comptime}
\end{figure}

\newpage
\section{Supplemental Tables}

\begin{table}[H]
\centering

\caption{Parameter estimates computed on data from Pleistocene epoch according to each of the three methods, displayed according to origin and terminus in the directed graph of $\hat{A}$. Parameters estimated as exactly zero by all three methods are excluded.}
\begin{tabular}{ll|lll}
  \hline
Origin & Terminus & Naive & Benchmark & Aggregation \\ 
  \hline
actinocyclus curvatulus & actinoptychus spp & 0.004 &  &  \\ 
  actinocyclus curvatulus & coscinodiscus spp & -0.002 &  &  \\ 
  actinocyclus curvatulus & roperia tesselata & -0.033 &  &  \\ 
  actinocyclus octonarius & actinocyclus octonarius & 0.017 &  &  \\ 
  actinocyclus octonarius & actinoptychus spp & 0.015 &  &  \\ 
  actinocyclus octonarius & coscinodiscus spp & -0.006 &  &  \\ 
  actinoptychus spp & actinocyclus curvatulus & 0.021 &  &  \\ 
  actinoptychus spp & actinocyclus octonarius & 0.011 &  &  \\ 
  actinoptychus spp & azpeitia nodulifer & -0.005 &  &  \\ 
  actinoptychus spp & cyclotella spp & -0.002 &  &  \\ 
  azpeitia nodulifer & actinocyclus curvatulus & -0.012 & -0.025 & -0.025 \\ 
  azpeitia nodulifer & actinoptychus spp & -0.016 & -0.017 & -0.017 \\ 
  azpeitia nodulifer & azpeitia nodulifer & 0.011 & 0.016 & 0.016 \\ 
  azpeitia nodulifer & cyclotella spp & 0.005 &  &  \\ 
  azpeitia nodulifer & roperia tesselata & -0.059 & -0.067 &  \\ 
  coscinodiscus spp & actinocyclus curvatulus & -0.018 &  &  \\ 
  coscinodiscus spp & actinocyclus octonarius & -0.027 &  &  \\ 
  coscinodiscus spp & actinoptychus spp & -0.012 &  &  \\ 
  coscinodiscus spp & azpeitia nodulifer & 0.011 &  &  \\ 
  coscinodiscus spp & coscinodiscus spp & 0.002 &  &  \\ 
  coscinodiscus spp & roperia tesselata & 0.010 &  &  \\ 
  cyclotella spp & actinocyclus octonarius & -0.029 &  &  \\ 
  cyclotella spp & cyclotella spp & 0.004 &  &  \\ 
  roperia tesselata & actinocyclus curvatulus & 0.021 &  &  \\ 
  roperia tesselata & actinoptychus spp & 0.011 &  &  \\ 
  roperia tesselata & azpeitia nodulifer & -0.056 &  &  \\ 
  roperia tesselata & coscinodiscus spp & 0.008 &  &  \\ 
  roperia tesselata & roperia tesselata & 0.034 &  &  \\ 
   \hline
\end{tabular}
\label{stbl:pleistocene-estimates}
\end{table}

\begin{table}[H]
\centering
\caption{Parameter estimates computed on data from Holocene epoch according to each of the three methods, displayed according to origin and terminus in the directed graph of $\hat{A}$. Parameters estimated as exactly zero by all three methods are excluded.}
\begin{tabular}{ll|lll}
  \hline
Origin & Terminus & Naive & Benchmark & Aggregation \\ 
  \hline
actinocyclus curvatulus & actinocyclus curvatulus & 0.008 &  &  \\ 
  actinocyclus curvatulus & roperia tesselata & -0.004 &  &  \\ 
  actinocyclus octonarius & cyclotella spp & -0.002 &  &  \\ 
  actinocyclus octonarius & roperia tesselata & 0.005 &  &  \\ 
  actinoptychus spp & actinocyclus octonarius & -0.004 &  &  \\ 
  actinoptychus spp & actinoptychus spp & 0.007 & 0.006 & 0.006 \\ 
  actinoptychus spp & azpeitia nodulifer & $|$a$|$$<$0.001 &  &  \\ 
  actinoptychus spp & cyclotella spp & -0.002 &  &  \\ 
  azpeitia nodulifer & actinocyclus octonarius & -0.006 &  &  \\ 
  azpeitia nodulifer & azpeitia nodulifer & 0.045 & 0.057 & 0.061 \\ 
  azpeitia nodulifer & coscinodiscus spp & -0.003 &  &  \\ 
  azpeitia nodulifer & cyclotella spp & 0.006 &  &  \\ 
  azpeitia nodulifer & roperia tesselata & -0.006 &  &  \\ 
  coscinodiscus spp & actinoptychus spp & -0.001 &  &  \\ 
  coscinodiscus spp & azpeitia nodulifer & -0.026 &  &  \\ 
  coscinodiscus spp & cyclotella spp & -0.003 &  &  \\ 
  coscinodiscus spp & roperia tesselata & 0.029 & 0.036 & 0.024 \\ 
  cyclotella spp & actinocyclus curvatulus & $|$a$|$$<$0.001 &  &  \\ 
  cyclotella spp & coscinodiscus spp & -0.006 & -0.007 & -0.008 \\ 
  cyclotella spp & cyclotella spp & 0.025 & 0.027 & 0.027 \\ 
  cyclotella spp & roperia tesselata & -0.008 & -0.006 &  \\ 
  roperia tesselata & actinocyclus curvatulus & -0.002 &  &  \\ 
  roperia tesselata & actinocyclus octonarius & 0.003 &  &  \\ 
  roperia tesselata & actinoptychus spp & 0.002 &  &  \\ 
  roperia tesselata & azpeitia nodulifer & -0.013 & -0.010 &  \\ 
  roperia tesselata & coscinodiscus spp & $|$a$|$$<$0.001 & 0.001 &  \\ 
  roperia tesselata & roperia tesselata & 0.010 & 0.012 &  \\ 
  roperia tesselata & stephanopyxis spp & -0.013 &  &  \\ 
   \hline
\end{tabular}
\label{stbl:holocene-estimates}
\end{table}

\end{appendices}

\end{document}

%% file: figures/fig-gvar-graph.tex
\begin{multicols}{2}

\centering
\begin{tikzpicture}
		    \tikzstyle{main} = [circle, 
		        minimum size = 12mm, 
		        line width = 0.4mm, 
		        draw=black!80,
		        node distance = 16mm]
		    \tikzstyle{empty} = [circle, 
		        minimum size = 12mm, 
		        line width = 0mm, 
		        draw=none,
		        node distance = 16mm]
		        
		    \node[main, fill = white!100] at (-3, 2) (1c) {$X_{t-2,1}$};
		    \node[main, fill = white!100] at (-3, 1) (2c) {$X_{t-2,2}$};
		    \node[empty, fill = none] at (-3, 0) (nc) {$\vdots$};
		    \node[main, fill = white!100] at (-3, -1.25) (Mc) {$X_{t-2,M}$};
		    \node[empty, fill = none] at (-3, -3) (cL) {$X_{t - 2}$};
		    
		    \node[main, fill = white!100] at (0, 2) (1a) {$X_{t-1,1}$};
		    \node[main, fill = white!100] at (0, 1) (2a) {$X_{t-1,2}$};
		    \node[empty, fill = none] at (0, 0) (na) {$\vdots$};
		    \node[main, fill = white!100] at (0, -1.25) (Ma) {$X_{t-1,M}$};
		    \node[empty, fill = none] at (0, -3) (aL) {$X_{t - 1}$};
		    
		    \node[main, fill = white!100] at (3, 2) (1b) {$X_{t,1}$};
		    \node[main, fill = white!100] at (3, 1) (2b) {$X_{t,2}$};
		    \node[empty, fill = none] at (3, 0) (nb) {$\vdots$};
		    \node[main, fill = white!100] at (3, -1.25) (Mb) {$X_{t,M}$};
		    \node[empty, fill = none] at (3, -3) (bL) {$X_t$};
		  		    
		    \node[empty, fill = none] at (-4.5, 2) (p1) {$\cdots$};
		    \node[empty, fill = none] at (-4.5, 1) (p2) {$\cdots$};
		    \node[empty, fill = none] at (-4.5, -1.25) (pM) {$\cdots$};
		    
		    \draw[->,  line width = 0.4mm] (1a) -- (2b) node[near start, above, sloped] {$a_{21}$};
		    \draw[->, line width = 0.4mm] (2a) -- (Mb) node[near start, above, sloped] {$a_{M2}$};
		    \draw[->, line width = 0.4mm] (Ma) -- (2b) node[near start, below, sloped] {$a_{2M}$};
		    
            \draw[->,  line width = 0.4mm] (1c) -- (2a) node[near start, above, sloped] {$a_{21}$};
		    \draw[->, line width = 0.4mm] (2c) -- (Ma) node[near start, above, sloped] {$a_{M2}$};
		    \draw[->, line width = 0.4mm] (Mc) -- (2a) node[near start, below, sloped] {$a_{2M}$};
		    
\end{tikzpicture}
\columnbreak
\null\vfill
\begin{equation*}
    A = \left(\begin{array}{cccc}
         {0} &{0} &\cdots &{0} \\
         a_{21} &{0} &\cdots &a_{2M} \\
         \vdots &\vdots &\ddots &\vdots \\
         {0} &a_{M2} &\cdots &{0} \\
    \end{array}\right)
\end{equation*}
\null\vfill
\end{multicols}


%% file: figures/fig-method-diagram.tex
\begin{tikzpicture}[node distance = 4.5cm]

\tikzstyle{data} = [rectangle split,
                rectangle split horizontal,
                rectangle split parts = 8,
                rectangle split part fill = {orange!30, orange!30!blue!20, white, red!30, red!30!blue!20, white, yellow!30, yellow!30!blue!20},
                minimum height = 1cm,
                thick,
                draw = black,
                align = center]
  
\tikzstyle{sample} = [rectangle,
                    minimum height = 1cm,
                    minimum width = 4.25cm,
                    thick,
                    draw = black,
                    align = center]

\tikzstyle{splitsample1} = [rectangle split, 
                      rectangle split horizontal,
                      rectangle split parts=2, 
                      thick,
                      draw=black, 
                      align=center,
                      minimum height = 1cm,
                      rectangle split part fill = {red!30, red!30!blue!30}]

\tikzstyle{splitsample2} = [rectangle split, 
                      rectangle split horizontal,
                      rectangle split parts=2, 
                      thick,
                      draw=black, 
                      align=center,
                      minimum height = 1cm,
                      rectangle split part fill = {orange!30, orange!30!blue!30}]
                      
\tikzstyle{splitsampleB} = [rectangle split, 
                      rectangle split horizontal,
                      rectangle split parts=2, 
                      thick,
                      draw=black, 
                      align=center,
                      minimum height = 1cm,
                      rectangle split part fill = {yellow!30, yellow!30!blue!30}]

\tikzstyle{emptyrec} = [rectangle, 
		        minimum height = 1cm, 
		        minimum width = 1cm,
		        line width = 0mm, 
		        draw=none,
		        text centered]
		        
\tikzstyle{outnode1} = [circle,
                thick,
                draw = black,
                align = center,
                fill = red!30]
                
\tikzstyle{outnode2} = [circle,
                thick,
                draw = black,
                align = center,
                fill = orange!30]
                
\tikzstyle{outnodeB} = [circle,
                thick,
                draw = black,
                align = center,
                fill = yellow!30]

 \node (data) [data, label = above:data $x$] {
        \nodepart[text width = 0.5cm]{one}
        \nodepart[text width = 0.25cm]{two}
        \nodepart[text width = 0.3cm]{three}
        \nodepart[text width = 0.5cm]{four}
        \nodepart[text width = 0.2cm]{five}
        \nodepart[text width = 1.5cm]{six}
        \nodepart[text width = 1cm]{seven}
        \nodepart[text width = 0.8cm]{eight}};

    \node (samp1) [splitsample1,
        below of = data, 
        yshift = 3cm,
        xshift = -5cm] {
        \nodepart[text width = 2cm]{one} $x^{*}_{11}$ train
        \nodepart[text width = 1.75cm]{two} $x^{*}_{12}$ test};

        \node (est11) [emptyrec, 
                below of = samp1, 
                yshift = 3.25cm, 
                xshift = -1.2cm,
                scale = 0.7] {
            $\hat{S}\left(\lambda_1; \textcolor{red!70}{x^{*}_{11}}\right)$};
 
        \node (mle11) [emptyrec, 
                right of = est11,
                xshift = -2.95cm,
                scale = 0.7] {
            $\hat{\beta}\left(\cdot\; ; \textcolor{red!70}{x^{*}_{11}}\right)$};

        \node (err11) [emptyrec, 
                right of = mle11,
                xshift = -3.02cm,
                scale = 0.7] {
            $e\left(\cdot\; ; \textcolor{red!70!blue!70}{x^{*}_{12}}\right)$};
            
        \draw [->, thick] ($(samp1.south west)!0.05!(samp1.south)$) |- (est11.west);
        \draw [->, thick] (est11) -- (mle11);
        \draw [->, thick] (mle11) -- (err11);
        
        \node (est12) [emptyrec, 
                below of = est11,
                yshift = 3.75cm,
                scale = 0.7] {
            $\hat{S}\left(\lambda_2; \textcolor{red!70}{x^{*}_{11}}\right)$};

        \node (mle12) [emptyrec, 
                right of = est12,
                xshift = -2.95cm,
                scale = 0.7] {
            $\hat{\beta}\left(\cdot\; ; \textcolor{red!70}{x^{*}_{11}}\right)$};
            
        \node (err12) [emptyrec, 
                right of = mle12,
                xshift = -3.02cm,
                scale = 0.7] {
            $e\left(\cdot\; ; \textcolor{red!70!blue!70}{x^{*}_{12}}\right)$};
 
        \draw [->, thick] ($(samp1.south west)!0.05!(samp1.south)$) |- (est12.west);
        \draw [->, thick] (est12) -- (mle12);
        \draw [->, thick] (mle12) -- (err12);
        
        \node (estdots1) [emptyrec, 
                below of = est12, 
                yshift = 4cm, 
                scale = 0.7] {
            $\vdots$};
        \node (mledots1) [emptyrec, 
                below of = mle12, 
                yshift = 4cm, 
                scale = 0.7] {
            $\vdots$};
        \node (errdots1) [emptyrec, 
                below of = err12, 
                yshift = 4cm, 
                scale = 0.7] {
            $\vdots$};
            
        \node (est1K) [emptyrec, 
                below of = estdots1,
                yshift = 3.75cm,
                scale = 0.7] {
            $\hat{S}\left(\lambda_K; \textcolor{red!70}{x^{*}_{11}}\right)$};

        \node (mle1K) [emptyrec, 
                right of = est1K,
                xshift = -2.95cm,
                scale = 0.7] {
            $\hat{\beta}\left(\cdot\; ; \textcolor{red!70}{x^{*}_{11}}\right)$};
            
        \node (err1K) [emptyrec, 
                right of = mle1K,
                xshift = -3.02cm,
                scale = 0.7] {
            $e\left(\cdot\; ; \textcolor{red!70!blue!70}{x^{*}_{12}}\right)$};
 
        \draw [->, thick] ($(samp1.south west)!0.05!(samp1.south)$) |- (est1K.west);
        \draw [->, thick] (est1K) -- (mle1K);
        \draw [->, thick] (mle1K) -- (err1K);
        
        \draw [decorate,
        decoration={brace, mirror, amplitude = 3pt, raise = 0pt},
        thick]
        (err1K.south west) -- (err1K.south east);
        
        \node (out1) [outnode1,
                below of = err1K,
                yshift = 3cm,
                xshift = 0cm,
                scale = 0.8] {
                $\hat{S}^*_1$};
                    
        \draw [->, thick] (err1K) + (0, -0.45cm) -- node[anchor = west, scale = 0.7] {min} (out1);
 
    \node (samp2) [splitsample2, right of = samp1, xshift = 0.2cm] {
        \nodepart[text width = 2cm]{one} $x^{*}_{21}$ train
        \nodepart[text width = 1.75cm]{two} $x^{*}_{22}$ test};
        
    \node (est21) [emptyrec, 
                below of = samp2, 
                yshift = 3.25cm, 
                xshift = -1.2cm,
                scale = 0.7] {
            $\hat{S}\left(\lambda_1; \textcolor{orange!70}{x^{*}_{21}}\right)$};
 
        \node (mle21) [emptyrec, 
                right of = est21,
                xshift = -2.95cm,
                scale = 0.7] {
            $\hat{\beta}\left(\cdot\; ; \textcolor{orange!70}{x^{*}_{21}}\right)$};

        \node (err21) [emptyrec, 
                right of = mle21,
                xshift = -3.02cm,
                scale = 0.7] {
            $e\left(\cdot\; ; \textcolor{orange!70!blue!70}{x^{*}_{22}}\right)$};
            
        \draw [->, thick] ($(samp2.south west)!0.05!(samp2.south)$) |- (est21.west);
        \draw [->, thick] (est21) -- (mle21);
        \draw [->, thick] (mle21) -- (err21);
        
        \node (est22) [emptyrec, 
                below of = est21,
                yshift = 3.75cm,
                scale = 0.7] {
            $\hat{S}\left(\lambda_2; \textcolor{orange!70}{x^{*}_{21}}\right)$};

        \node (mle22) [emptyrec, 
                right of = est22,
                xshift = -2.95cm,
                scale = 0.7] {
            $\hat{\beta}\left(\cdot\; ; \textcolor{orange!70}{x^{*}_{21}}\right)$};
            
        \node (err22) [emptyrec, 
                right of = mle22,
                xshift = -3.02cm,
                scale = 0.7] {
            $e\left(\cdot\; ; \textcolor{orange!70!blue!70}{x^{*}_{22}}\right)$};
 
        \draw [->, thick] ($(samp2.south west)!0.05!(samp2.south)$) |- (est22.west);
        \draw [->, thick] (est22) -- (mle22);
        \draw [->, thick] (mle22) -- (err22);
        
        \node (estdots2) [emptyrec, 
                below of = est22, 
                yshift = 4cm, 
                scale = 0.7] {
            $\vdots$};
        \node (mledots2) [emptyrec, 
                below of = mle22, 
                yshift = 4cm, 
                scale = 0.7] {
            $\vdots$};
        \node (errdots2) [emptyrec, 
                below of = err22, 
                yshift = 4cm, 
                scale = 0.7] {
            $\vdots$};
            
        \node (est2K) [emptyrec, 
                below of = estdots2,
                yshift = 3.75cm,
                scale = 0.7] {
            $\hat{S}\left(\lambda_K; \textcolor{orange!70}{x^{*}_{21}}\right)$};

        \node (mle2K) [emptyrec, 
                right of = est2K,
                xshift = -2.95cm,
                scale = 0.7] {
            $\hat{\beta}\left(\cdot\; ; \textcolor{orange!70}{x^{*}_{21}}\right)$};
            
        \node (err2K) [emptyrec, 
                right of = mle2K,
                xshift = -3.02cm,
                scale = 0.7] {
            $e\left(\cdot\; ; \textcolor{orange!70!blue!70}{x^{*}_{22}}\right)$};
 
        \draw [->, thick] ($(samp2.south west)!0.05!(samp2.south)$) |- (est2K.west);
        \draw [->, thick] (est2K) -- (mle2K);
        \draw [->, thick] (mle2K) -- (err2K);
        
        \draw [decorate,
        decoration={brace, mirror, amplitude = 3pt, raise = 0pt},
        thick]
        (err2K.south west) -- (err2K.south east);
        
        \node (out2) [outnode2,
                below of = err2K,
                yshift = 3cm,
                xshift = 0cm,
                scale = 0.8] {
                $\hat{S}^*_2$};
                    
        \draw [->, thick] (err2K) + (0, -0.45cm) -- node[anchor = west, scale = 0.7] {min} (out2);

    \node (dots) [emptyrec, right of = samp2, xshift = -1.85cm] {$\cdots$};
    
    \node (sampB) [splitsampleB, right of = dots, xshift = -1.9cm] {
        \nodepart[text width = 2cm]{one} $x^{*}_{J1}$ train
        \nodepart[text width = 1.75cm]{two} $x^{*}_{J2}$ test};
        
    \node (estB1) [emptyrec, 
                below of = sampB, 
                yshift = 3.25cm, 
                xshift = -1.2cm,
                scale = 0.7] {
            $\hat{S}\left(\lambda_1; \textcolor{yellow!70!black!60}{x^{*}_{J1}}\right)$};
 
        \node (mleB1) [emptyrec, 
                right of = estB1,
                xshift = -2.95cm,
                scale = 0.7] {
            $\hat{\beta}\left(\cdot\; ; \textcolor{yellow!70!black!60}{x^{*}_{J1}}\right)$};

        \node (errB1) [emptyrec, 
                right of = mleB1,
                xshift = -3.02cm,
                scale = 0.7] {
            $e\left(\cdot\; ; \textcolor{yellow!70!blue!70}{x^{*}_{J2}}\right)$};
            
        \draw [->, thick] ($(sampB.south west)!0.05!(sampB.south)$) |- (estB1.west);
        \draw [->, thick] (estB1) -- (mleB1);
        \draw [->, thick] (mleB1) -- (errB1);
        
        \node (estB2) [emptyrec, 
                below of = estB1,
                yshift = 3.75cm,
                scale = 0.7] {
            $\hat{S}\left(\lambda_2; \textcolor{yellow!70!black!60}{x^{*}_{J1}}\right)$};

        \node (mleB2) [emptyrec, 
                right of = estB2,
                xshift = -2.95cm,
                scale = 0.7] {
            $\hat{\beta}\left(\cdot\; ; \textcolor{yellow!70!black!60}{x^{*}_{J1}}\right)$};
            
        \node (errB2) [emptyrec, 
                right of = mleB2,
                xshift = -3.02cm,
                scale = 0.7] {
            $e\left(\cdot\; ; \textcolor{yellow!70!blue!70}{x^{*}_{J2}}\right)$};
 
        \draw [->, thick] ($(sampB.south west)!0.05!(sampB.south)$) |- (estB2.west);
        \draw [->, thick] (estB2) -- (mleB2);
        \draw [->, thick] (mleB2) -- (errB2);
        
        \node (estdotsB) [emptyrec, 
                below of = estB2, 
                yshift = 4cm, 
                scale = 0.7] {
            $\vdots$};
        \node (mledotsB) [emptyrec, 
                below of = mleB2, 
                yshift = 4cm, 
                scale = 0.7] {
            $\vdots$};
        \node (errdotsB) [emptyrec, 
                below of = errB2, 
                yshift = 4cm, 
                scale = 0.7] {
            $\vdots$};
            
        \node (estBK) [emptyrec, 
                below of = estdotsB,
                yshift = 3.75cm,
                scale = 0.7] {
            $\hat{S}\left(\lambda_K; \textcolor{yellow!70!black!60}{x^{*}_{J1}}\right)$};

        \node (mleBK) [emptyrec, 
                right of = estBK,
                xshift = -2.95cm,
                scale = 0.7] {
            $\hat{\beta}\left(\cdot\; ; \textcolor{yellow!70!black!60}{x^{*}_{J1}}\right)$};
            
        \node (errBK) [emptyrec, 
                right of = mleBK,
                xshift = -3.02cm,
                scale = 0.7] {
            $e\left(\cdot\; ; \textcolor{yellow!70!blue!70}{x^{*}_{J2}}\right)$};
 
        \draw [->, thick] ($(sampB.south west)!0.05!(sampB.south)$) |- (estBK.west);
        \draw [->, thick] (estBK) -- (mleBK);
        \draw [->, thick] (mleBK) -- (errBK);
        
        \draw [decorate,
        decoration={brace, mirror, amplitude = 3pt, raise = 0pt},
        thick]
        (errBK.south west) -- (errBK.south east);
        
        \node (outB) [outnodeB,
                below of = errBK,
                yshift = 3cm,
                xshift = 0cm,
                scale = 0.8] {
                $\hat{S}^*_J$};
                    
        \draw [->, thick] (errBK) + (0, -0.45cm) -- node[anchor = west, scale = 0.7] {min} (outB);
        
    \node (out) [emptyrec,
        below of = out2,
        yshift = 3.25cm,
        xshift = 0cm] {
        $\hat{S}^* = \left\{i: 
        \frac{1}{J}\sum_{j = 1}^J \mathbf{1}\{i \in \hat{S}^*_j \} \geq \gamma
        \right\}$
        };
	
	\draw [->, thick] (out1) -- (out.north west);
	\draw [->, thick] (out2) -- (out.north);
	\draw [->, thick] (outB) -- (out.north east);
\end{tikzpicture}
